\documentclass[%
superscriptaddress,
preprint,
amsmath,amssymb,
aps, 
physrev,
]{revtex4-2}

\usepackage{amsmath}
\usepackage{amssymb}
\usepackage{xfrac}
\usepackage{graphicx}
\usepackage{caption}
\usepackage{subcaption}
\graphicspath{ {images/} }
\usepackage[dvipsnames]{xcolor}
\usepackage[margin=1in]{geometry}
\usepackage{soul}
\usepackage{comment}
\usepackage{float}
\usepackage{color}
\usepackage{xcolor}
\usepackage{etoolbox}
\usepackage{parskip}
\usepackage{microtype}
\usepackage{mathtools}
\usepackage{float}
\usepackage{tikz, pgfplots}
\usepackage[scr=boondox]{mathalpha} 
\usepackage[thinc]{esdiff}

\usetikzlibrary{arrows.meta}
\usetikzlibrary{positioning}
\usetikzlibrary{calc}
\usetikzlibrary{backgrounds}

\definecolor{main}{RGB}{  0,   0,   0}
\definecolor{light1}{HTML}{4e79a7}
\definecolor{light2}{HTML}{f28e2b}
\definecolor{light3}{HTML}{59a14f}
\definecolor{light4}{HTML}{e15759}
\definecolor{light5}{HTML}{76b7b2}
\definecolor{light6}{HTML}{b07aa1}
\definecolor{light7}{HTML}{ff9da7}
\definecolor{light8}{HTML}{76b7b2}
\colorlet{dark1}{light1!70!main}
\colorlet{dark2}{light2!70!main}
\colorlet{dark3}{light3!70!main}
\colorlet{dark4}{light4!70!main}
\colorlet{dark5}{light5!70!main}

\definecolor{lightblue}{rgb}{0.63, 0.74, 0.78}
\definecolor{seagreen}{rgb}{0.18, 0.42, 0.41}
\definecolor{orange}{rgb}{0.85, 0.55, 0.13}
\definecolor{silver}{rgb}{0.69, 0.67, 0.66}
\definecolor{rust}{rgb}{0.72, 0.26, 0.06}
\definecolor{purp}{RGB}{68, 14, 156}
\definecolor{joshua}{RGB}{251,220,127}
\definecolor{sky}{RGB}{86,180,233}

\colorlet{lightsilver}{silver!30!white}
\colorlet{darkorange}{orange!75!black}
\colorlet{darksilver}{silver!65!black}
\colorlet{darklightblue}{lightblue!70!black}
\colorlet{darkrust}{rust!85!black}
\colorlet{darkseagreen}{seagreen!85!black}
\definecolor{darksky}{HTML}{154c79}

\newcommand{\ens}[1]{\langle #1 \rangle}

\newcommand{\bigpar}[1]{\left( #1 \right)}
\newcommand{\bigbra}[1]{\left[ #1 \right]}
\newcommand{\tsf}{\ens{-\rho v''Y_H''}}

\newcommand{\rhom}{\ens{\rho}}
\newcommand{\YHm}{\widetilde{Y_H}}
\newcommand{\Ycm}{\widetilde{Y_c}}

\newcommand{\vm}{\widetilde{v}}
\newcommand{\DDt}[1]{\frac{D#1}{Dt}}
\newcommand{\cpp}[1]{\mathscr{c}^{#1}}
\newcommand{\diffu}[1]{\diffp{}{{x_j}}\rho D_H\diffp{}{{x_j}}{#1}}

\newcommand{\cmom}{\mathscr{c}}

\newcommand{\edit}[1]{\textcolor{black}{#1}}
\newcommand{\editt}[1]{\textcolor{black}{#1}}

\usepackage[colorlinks=true,linkcolor=darkrust,citecolor=darklightblue,urlcolor=darksilver]{hyperref}
\usepackage{cleveref}

\expandafter\def\csname ver@etex.sty\endcsname{}

\usepackage[font=small,format=plain,labelfont=bf]{caption}

\crefname{appendix}{}{}

\usepackage{fancyhdr}

\begin{document}
	

\hypersetup{
  linkcolor=darkrust,
  citecolor=seagreen,
  urlcolor=darkrust,
  pdfauthor=author,
}

%
%
%
%

\preprint{APS/123-QED}

\title{\textbf{Atwood effects on nonlocality of the scalar transport closure in Rayleigh-Taylor mixing} 
}%

\author{Dana L. O.-L. Lavacot}
\email{Contact author: dlol@stanford.edu}
\affiliation{%
	Department of Mechanical Engineering, Stanford University, California 94305, USA
}%

\author{Ali Mani}
\affiliation{%
	Department of Mechanical Engineering, Stanford University, California 94305, USA
}%

\author{Brandon E. Morgan}
\affiliation{
	Lawrence Livermore National Laboratory, California 94550, USA
}%

\date{\today}

\begin{abstract}
The importance of nonlocality is assessed in modeling mean scalar transport for \edit{turbulent} Rayleigh-Taylor (RT) mixing \edit{at different Atwood numbers}.
Building on the two-dimensional incompressible work of \citet{lavacot2024}, the present work extends the Macroscopic Forcing Method (MFM) to variable density problems in three-dimensional space to measure moments of the generalized eddy diffusivity kernel in RT mixing for increasing Atwood numbers ($A=0.05, 0.3, 0.5, 0.8$).
It is found that as $A$ increases: 1) the eddy diffusivity moments become asymmetric, and 2) the higher-order eddy diffusivity moments become larger relative to the leading-order diffusivity, indicating that nonlocality becomes more important at higher $A$.
There is a particularly strong temporal nonlocality at higher $A$, suggesting stronger history effects.
The implications of these findings for closure modeling for finite-Atwood RT are discussed.


\end{abstract}

\maketitle


\section{Introduction}
Rayleigh-Taylor (RT) instability occurs when a light fluid is accelerated into a heavy fluid through a perturbed interface.
Over time, the instability results in turbulent mixing and enters a self-similar regime.
Understanding the effects of this mixing is critical in engineering design applications, especially for inertial confinement fusion (ICF).
In ICF, a plastic ablator is accelerated into deuterium gas to achieve high compression and, consequently, ignition.
If perterbations exist at material interfaces in the capsule, RT instability will cause these perturbations to grow, which may result in mixing that ultimately reduces energy output \citep{pak2020, zhou2025instabilities}. 


In designing experiments for ICF, it is crucial to accurately predict the turbulent mixing in simulations.
High-fidelity approaches such as direct numerical simulations and large eddy simulations have been used to accurately predict turbulent mixing in RT instability \citep{youngs1994, cabotcook2006, mueschkeschilling2009}.
However, the fine grids required to resolve turbulent scales make these methods prohibitively expensive for the iterative design process, in which thousands of simulations must be run.
A more computationally feasible approach is simulation of the Reynolds-Averaged Navier-Stokes (RANS) equations, which requires resolution of only the ensemble-averaged fields.
RANS has been previously employed in optimizing designs for ICF experiments \citep{dittrich2014design, casey2014, sterbentz2022design, weber2023reduced}.
Since only the mean quantities are evolved in RANS simulations, models are required to approximate the mean impact of the underlying fluctuations.
It is then crucial in the RANS approach to use accurate models for the unclosed terms.

The present work focuses on scalar transport, which characterizes the mixing of materials in RT instability.
RANS modeling of scalar transport involves closing the turbulent species flux in the mean scalar transport equation.
A common closure for this term is a gradient-diffusion approximation, which assumes the flux depends only on local gradients of the mean scalar field.
A gradient diffusion closure is used in popular models for RT mixing such as the $k$--$\epsilon$ model \citep{launder1974} (modified by \citet{gauthierbonnet1990} for RT) and the $k$--$L$ model \citep{dimontetipton2006}.
However, the gradient-diffusion closure has been shown to be insufficient for modeling scalar transport in RT mixing \citep{morgan2017, denissen2014}.
Alternative models include those that use transport equations for the flux itself \citep{braungore2021} and those based on two-point correlations \citep{clark1995, steinkamp_i_1999, pal2018}, the latter of which are intended to capture nonlocality.
While these models address nonlocal effects in RT mixing, they do so without directly examining the nonlocality of the closure operator.
Traditional approaches towards studying nonlocality usually involve examination of two-point correlations.

Nonlocality of the eddy diffusivity has been studied in a previous work \citep{lavacot2024} for  two-dimensional (2D) RT mixing in the Boussinesq limit ($A=0.05$).
In that work, the Macroscopic Forcing Method (MFM) \citep{manipark2021} was used to measure moments of the generalized eddy diffusivity kernel, which describes the nonlocal dependence of the turbulent flux on mean gradients.
MFM is similar to the Green's function approach described by \citet{hamba2005} for determining the exact nonlocal eddy diffusivity but also allows for polynomial forcings that enable measurement of eddy diffusivity moments.
Measuring the moments is more efficient than computing the full kernel, which becomes expensive for unsteady problems with large macroscopic spaces, like RT mixing.

The goal of the present work is to extend MFM analysis to turbulent three-dimensional (3D) RT mixing at higher Atwood numbers.
The extension to 3D is certainly necessary for investigation of truly turbulent RT.
Additionally, simulations of higher Atwood numbers allow for the investigation of variable density effects on nonlocality.
It is known that the behavior of the instability differs between the variable density regime and the Boussinesq limit.
Particularly, asymmetry in the mixing layer arises at higher Atwood numbers, which also leads to asymmetries in turbulent statistics, as shown by \citet{livescu2010}.
Thus, one purpose of this study is to investigate the strength of the dependence of nonlocality in RT mixing on Atwood number.
Ultimately, findings from this study will inform development of more accurate turbulence models for variable density RT mixing that incorporate nonlocality and its dependence on Atwood number.

This work is organized as follows.
Numerical methods of this work are presented in \S \ref{sec:methods}, in which the governing equations and their numerical solution are described along with a brief overview of RT physics.
In \S \ref{sec:nonlocality}, moments of the spatio-temporally nonlocal eddy diffusivity are measured using MFM and analyzed.
The importance of the eddy diffusivity moments in modeling is examined in \S \ref{sec:modeling}.
Finally, the results and implications for modeling are discussed in the Conclusion in \S \ref{sec:conclusion}.

\section{Numerical methods}
\label{sec:methods}
\subsection{Governing equations}
\label{subsec:goveq}
\edit{
An RT configuration is considered in which a heavy fluid sits on top of a light fluid, and gravity points in the negative $y$-direction.
	The fluids are miscible, and conservation of mass gives that the mass fraction of the heavy fluid, $Y_H$, and the light fluid, $Y_L$, sum to unity.
The flow} is governed by the compressible Navier-Stokes equations:

\begin{align}
    \frac{\partial \rho Y_k}{\partial t}&=-\frac{\partial }{\partial x_j} \bigpar{ \rho u_j Y_k - \rho D_k\frac{\partial Y_k}{\partial x_j} },\label{eq:scalar_transport}\\
    \frac{\partial \rho u_i}{\partial t}&=-\frac{\partial }{\partial x_j} \bigpar{ \rho u_i u_j + p\delta_{ij} - \tau_{ij} } + {\rho g_i,}\\
    \frac{\partial E}{\partial t}&=-\frac{\partial }{\partial x_j} \bigpar{ \bigpar{E+p}u_j - \tau_{ij}u_i + q_i } + \rho g_i u_i.\label{eq:NS_end}
\end{align}
Here, $\rho$ is density, $u_i$ is velocity, $Y_k$ is the mass fraction of species $\alpha$, $D_k$ is the diffusivity of species $\alpha$, $p$ is pressure, $\delta_{ij}$ is the Kronecker delta, $\tau_{ij}$ is the viscous stress, $g_i$ is gravitational acceleration (in this work, $\mathbf{g}=(0,-g,0)^T$), $E=\rho\bigpar{e+\frac{1}{2}u_ku_k}$ is the total energy, and $q_i$ is the thermal energy flux.
The viscous stress is
\begin{align}
    \tau_{ij} = \mu\bigpar{\diffp{u_i}{{x_j}}+\diffp{u_j}{{x_i}}} + \bigpar{\beta-\frac{2}{3}\mu} \frac{\partial u_k}{\partial x_k} \delta_{ij},
\end{align}
where $\mu$ is the molecular viscosity and $\beta$ is the bulk viscosity.
The energy flux is 
\begin{align}
    q_i = -\kappa \diffp{T}{{x_i}},    
\end{align}
where $T$ is temperature.
Pressure and temperature are determined using the ideal gas law:
\begin{align}
    p = \rho\bigpar{\gamma-1}e, \quad T = \bigpar{\gamma-1}\frac{e}{R},
\end{align}
where $\gamma$ is the ratio of specific heats $\frac{c_p}{c_v}$ and $R$ is the specific constant.
\edit{
The specific heats are dependent on $Y_k$:
\begin{align}
	c_v=Y_H c_{v,H} + Y_L c_{v,L},\quad
	c_p=Y_H c_{p,H} + Y_L c_{p,L},
\end{align}
}
\edit{
In general, the energy equation (Equation \ref{eq:NS_end}) would include an enthalpy diffusion term to account for energy fluxes due to the species' diffusion, as described by \citet{cook2009}.
However, since $\Delta T$ across the heavy-light fluid interface at the $A$ studied in this work is expected to be small, we choose to neglect this term.
Additionally, our analysis is in the late-time self-similar regime, where the enthalpy diffusion term is expected to be small.
}

\edit{
For MFM analysis, we also consider transport of a dilute scalar $Y_c$, which is governed by the following transport equation:
\begin{align}
	\frac{\partial \rho Y_c}{\partial t}&=-\frac{\partial }{\partial x_j} \bigpar{ \rho u_j Y_c - \rho D_c\frac{\partial Y_c}{\partial x_j} }.\label{eq:ste_dilute}
\end{align}
For this dilute scalar, coupling from momentum is one way, and $Y_c$ does not affect $\rho$.
Thus, the transport equation for $Y_c$ is linear.
}


\subsection{RT mixing and self-similarity}
\label{subsec:selfsim}
The difference in density between the two fluids can be expressed nondimensionally as the Atwood number:
\begin{align}
    A=\frac{\rho_H-\rho_L}{\rho_H+\rho_L},
\end{align}
where $\rho_H$ is the density of the heavy fluid, and $\rho_L$ is the density of the light fluid.

As the instability develops, bubbles rise into the heavy fluid, and spikes sink into the light fluid.
Over time, secondary Kelvin-Helmholtz instabilities are triggered, and the flow transitions into turbulence.
In this turbulent state, RT instability becomes self-similar, and the growth of the bubbles and spikes are quadratic in time \citep{youngs1984numerical}:
\begin{align}
    h_b\approx\alpha_b Agt^2, \quad
     h_s\approx-\alpha_s Agt^2,
\end{align}
where $h_b$ and $h_s$ are the bubble and spike heights, respectively, and $\alpha_b$ and $\alpha_s$ are the bubble and spike growth parameters, respectively.
Based on the bubble height, a self-similar variable in space is defined:
\begin{align}
	\eta\equiv\frac{{y-\frac{1}{2}}}{h_b},
\end{align} 
for $y$ defined between $0$ and $1$.
The mixing layer half-width is defined as $h=\frac{1}{2}\bigpar{h_b - h_s}$, and in the self-similar limit, the growth of $h$ can be characterized with a single $\alpha$:
\begin{align}
    h\approx\alpha Agt^2.
\end{align}
At low $A$, $h_b\approx h_s$.
Increasing $A$ increases the asymmetry of the mixing layer \citep{youngs2013, livescu2010} as the spikes sink faster than the bubbles rise.
Thus, for finite $A$, $h>h_b$.
In this work, the growth of the bubbles is used for self-similar analysis rather than the total mixing layer growth parameter $\alpha$, since the latter varies significantly across Atwood numbers \citep{zhou2017a}.

Using the analytical derivation of the mixing width from \citet{ristorcelliclark2004}, \citet{cabotcook2006} defines the bubble growth parameter as
\begin{align}
    \alpha_b = \frac{\dot{h_b}^2}{4Agh_b},\label{eq:alpha_b}
\end{align}
where $\dot{h_b}$ is the rate of change of $h_b$ in time.
This definition is used to observe the growth parameter over time and assess convergence to self-similarity.
In the self-similar regime, $\alpha_b$ should converge to a constant value over time.

For self-similar analysis, the definition of $\alpha_b$ by \citet{livescu2010} is used:
\begin{align}
	\alpha_b = \left(\frac{h_b(t)^{1/2}-h_b(t_0)^{1/2}}{(Ag)^{1/2}(t-t_0)}\right)^2
	\label{eq:alpha_b_est}
\end{align}
where $t_0$ is an arbitrary time during the self-similar growth of the mixing layer.
This definition is preferable for self-similar fits and normalizations, since it avoids temporal derivatives of the mixing width, which is not smooth in time due to statistical error.




The bubble height can be computed from mass fraction profiles by taking it as the distance from the centerline of the domain to where the mean mass fraction of the light fluid is $0.999$.
The RT instability can be considered self-similar when this $h_{b,99}$ becomes quadratic in time.

Another metric for self-similarity is the mixedness parameter, defined as
\begin{align}
    \phi \equiv 1-4\frac{\int\widetilde{ Y_H''Y_H''} dy}{\int\widetilde{ Y_H}\widetilde{Y_L} dy},\label{eq:phi}
\end{align}
where $Y_H$ is the mass fraction of the heavy fluid, and $Y_L=1-Y_H$ is the mass fraction of the light fluid.
For self-similar RT mixing, $\phi$ is expected to converge to a steady-state value of about 0.8 \cite{cabotcook2006}.

\edit{
	It is additionally useful to assess the development of the RT flow by examining relevant Reynolds numbers.
}
The Taylor microscale Reynolds number is defined as
\begin{align}
    Re_T = \frac{k^{1/2}\lambda}{\nu},
\end{align}
where $k=\frac{1}{2}\widetilde{u_i''u_i''}$, and
\begin{align}
    \lambda = \sqrt{\frac{10\nu L}{k^{1/2}}}.
    \label{eq:ReT}
\end{align}
$L$ is a turbulent length scale, which can be approximated as $\frac{1}{5}$ of the total mixing width, $h_b+h_s$ \citep{morgan2017}.
The large scale Reynolds number \citep{cabotcook2006} is defined as
\begin{align}
    Re_L = \frac{h_{99}\dot{h}_{99}}{\nu},
\end{align}
where $h_{99}$ is the total mixing width defined as the distance between the locations of mass fractions $0.001$ and $0.999$.

\edit{
	\citet{dimotakis2000} describes a ``mixing transition,'' after which the amount of mixed fluid depends more weakly on Reynolds number.
	This mixing transition was identified to be when $Re_T>100$ and $Re_L>10,000$; we refer to these conditions here as critical Reynolds numbers for mixing.
	\citet{zhou2007} also identifies a minimum Reynolds number for turbulence, $Re_L>1.6\times 10^5$; we refer to this as the critical Reynolds number for turbulence.
	While both of these transitions occur before the transition to the self-similar regime, they are still useful metrics for understanding the stage of development of an RT flow.
}

\subsection{Numerical solution to governing equations}
\label{subsec:numsol}
\textit{Pyranda} \cite{olson2023}, an open source finite difference solver, is used to solve Equations \ref{eq:scalar_transport}-\ref{eq:NS_end}.
Its numerical methods are the same as those used in \textit{Miranda}, a hydrodynamics code developed at Lawrence Livermore National Laboratory \citep{schulz2010, rehagen2017}.
The codes use fourth-order Runge-Kutta in time and a tenth-order compact differencing scheme in space, and due to this high-order spatial scheme, they use artificial fluid properties for stability.
Details on the artificial fluid method can be found in Appendix \ref{appendix:artificial_fluid}

To prevent the numerical diffusion from dominating the physical turbulent diffusion, the numerical Grashof number is kept small.
This Grashof number is defined as
\begin{align}
    Gr = \frac{-2gA\Delta^3}{\nu^2}.
\end{align}
where $\Delta=\Delta_x=\Delta_y=\Delta_z$ is the grid spacing.
In line with the findings of \citet{morganblack2020}, $Gr=12$ is used in order to keep numerical diffusion finite but still allow turbulence to develop before the edges of the mixing layer reach the domain boundaries. 
\edit{
	The authors also approximate the ratio of the Kolmogorov scale $\eta_K$ to grid size $\Delta$ as 
	\begin{align}
		\frac{\eta_K}{\Delta} \approx \bigpar{\frac{12^3}{Gr^3 N_h}}^{\frac{1}{8}},
	\end{align}
	where $N_h$ is the number of points across the mixing layer, $\frac{h}{\Delta}$.
	Based on this, $\eta_K/\Delta\approx 0.5$ at the end of the simulations, indicating that resolution extends into the viscous range.
}

\begin{table}
    \centering
    \begin{tabular}{c|c|c|c|c}
         Nondimensional number & $A=0.05$  & $A=0.3$  & $A=0.5$  & $A=0.8$ \\
         \hline
         $Ma_\text{max}$ & $0.04 $ & $  0.12 $ & $0.17 $ & $0.32 $\\
         $Re_T, Pe_T$ & $114  $ & $  112  $ & $113 $ & $114 $\\
         $Re_L, Pe_L$ & $1.34\times 10^4 $ & $  1.15 \times 10^4  $ & $1.27\times 10^4 $ & $1.23\times 10^4 $
    \end{tabular}
    \caption{Values of nondimensional numbers at the end of the simulation for each $A$ case.
    $Ma_\text{max}$ is reported from one realization.
    The Reynolds and Peclet numbers are computed from averaged realizations.}
    \label{tab:nondim}
\end{table}

The other relevant nondimensional numbers of the RT mixing problem are the Mach number ($Ma$), Peclet number ($Pe$), and Schmidt number ($Sc$):
\begin{align}
    Ma &= \frac{u}{c},\\
    Pe_T &= Re_T Sc,\\
    Pe_L &= Re_L Sc,\\
    Sc &= \frac{\nu}{D_M}.
\end{align}
Here, $c$ is the speed of sound and is set by the heat capacity ratio $\gamma$, which is $5/3$ in the simulations presented here.
The Schmidt number is chosen to be unity.
The Peclet numbers are determined by the Reynolds and Schmidt numbers, and the Reynolds numbers are set through the numerical Grashof number, which fixes $\nu$ through choice of $g$, $A$, and $\Delta$.
Values of these nondimensional numbers at the final timesteps of the simulations are listed in Table \ref{tab:nondim}.


The simulation domain is $0.5\times1.0\times0.5$ cm, so the length in $y$ is twice the lengths in $x$ and $z$, and the grid is $512\times1024\times512$ cells.
The domain is periodic in $x$ and $z$, and no slip and no penetration conditions are applied at $y=0$ and $y=1$ cm.
Sponge layers of finite thickness are applied to the velocity and density fields at the boundaries in $y$ to prevent reflection of acoustic waves that arise from the high-order numerics. 

The mass fraction profile is initialized as a $\tanh$ profile with approximately ten cells across the interface.
A multi-mode perturbation is added at the interface:
\begin{align}
    E_0 =& \frac{\Delta/2}{\kappa_\text{max}-\kappa_\text{min}+1},\\
    \xi(x,z) =&
    E_0
    \sum^{\kappa_\text{max}}_{k_x=\kappa_\text{min}} \sum^{\kappa_\text{max}}_{k_z=\kappa_\text{min}}
   \bigpar{\cos\bigpar{2\pi k_x x+\phi_{1,k}}
    +\sin\bigpar{\edit{2}\pi k_x x+\phi_{2,k}}}\nonumber
    \\&
     \quad\quad\quad\quad\quad\quad\times  \bigpar{\cos\bigpar{2\pi k_z z+\phi_{3,k}}
    +\sin\bigpar{\edit{2}\pi k_z z+\phi_{4,k}}},\\
    Y_H(x,y,z) =& \frac{1}{2}\left(1+\tanh\left(\frac{y-L_y/2-\xi}{2\Delta}\right)\right),
\end{align}
where $\phi_{1,k}$, $\phi_{2,k}$ , $\phi_{3,k}$ , and $\phi_{4,k}$ are phase shift vectors randomly taken from a uniform distribution.
The minimum and maximum wavenumbers are set to  $\kappa_\text{min}=8$ and $\kappa_\text{max}=64$, respectively.
\edit{
	These minimum and maximum wavenumbers are within the range found by \citet{livescu2011} to result in converged $\alpha$ by the end of their DNS of RT instability.
}

Density is computed from this initial mass fraction profile as $\rho_H Y_H + \rho_L Y_L$.
The light fluid density is set to unity for all simulations, and the heavy fluid density is determined from this and the Atwood number.
Pressure is initialized as a hydrostatic pressure based on the initial density field, $p=\rho g \bigpar{y-\frac{1}{2}} + 1$.
The velocity field \edit{is} initially zero.

The simulation is stopped when $h_{99}$ reaches $0.5$ cm\edit{, which is half the computational domain in $y$.
	This stop condition is chosen to avoid the effects of lateral confinement \citep{thevenin2025} as well as allow }the mixing layer to \edit{reach the turbulent and self-similar regimes} while avoiding interference from the top and bottom walls.


\begin{figure}
    \centering
    \begin{subfigure}[]{0.35\textwidth}
        \includegraphics[width=\textwidth]{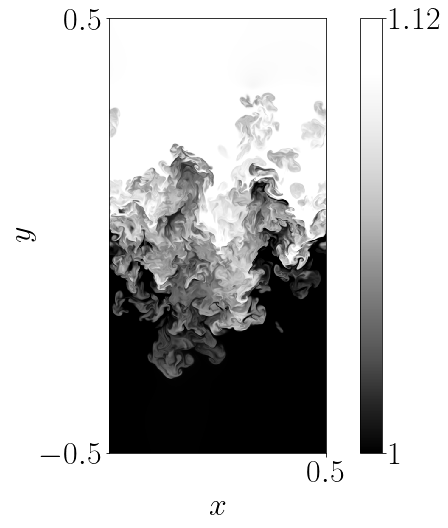}
        \subcaption[]{$A=0.05$}
    \end{subfigure}
    \begin{subfigure}[]{0.35\textwidth}
        \includegraphics[width=\textwidth]{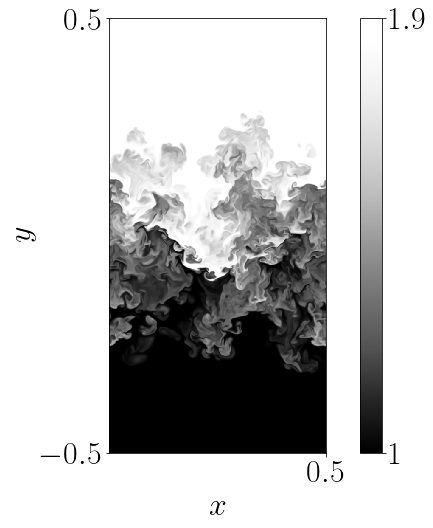}
        \subcaption[]{$A=0.3$}
    \end{subfigure}\\
    \begin{subfigure}[]{0.35\textwidth}
        \includegraphics[width=\textwidth]{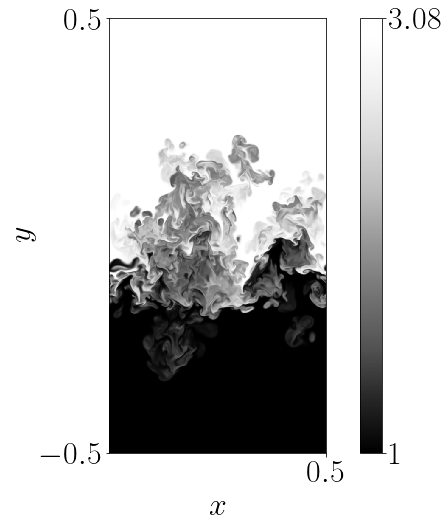}
        \subcaption[]{$A=0.5$}
    \end{subfigure}
    \begin{subfigure}[]{0.35\textwidth}
        \includegraphics[width=\textwidth]{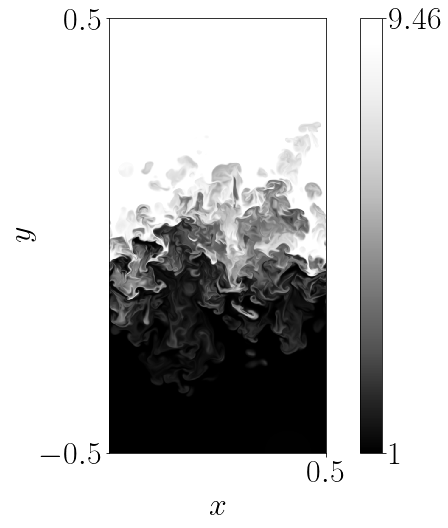}
        \subcaption[]{$A=0.8$}
    \end{subfigure}
    \caption{Contours of density for each $A$ case.}
    \label{fig:contours_rho}
\end{figure}
\begin{figure}
    \centering
    \begin{subfigure}[]{0.35\textwidth}
        \includegraphics[width=\textwidth]{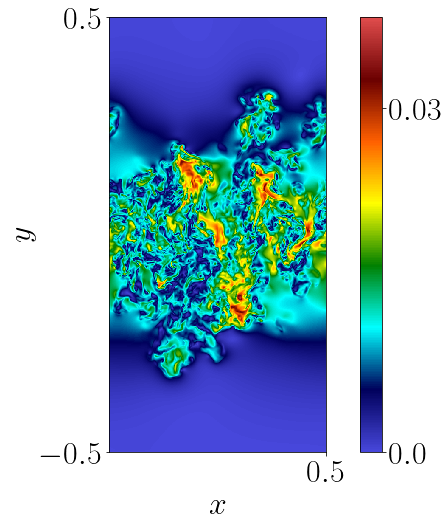}
        \subcaption[]{$A=0.05$}
    \end{subfigure}
    \begin{subfigure}[]{0.35\textwidth}
        \includegraphics[width=\textwidth]{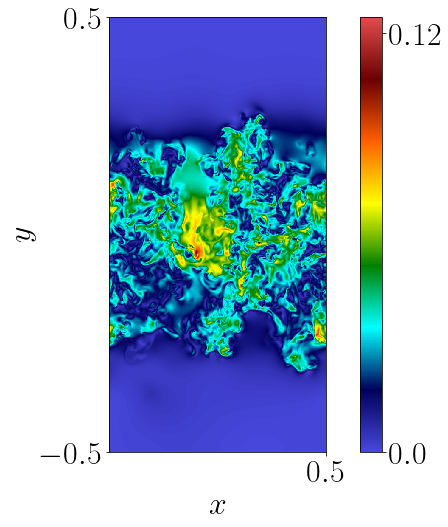}
        \subcaption[]{$A=0.3$}
    \end{subfigure}\\
    \begin{subfigure}[]{0.35\textwidth}
        \includegraphics[width=\textwidth]{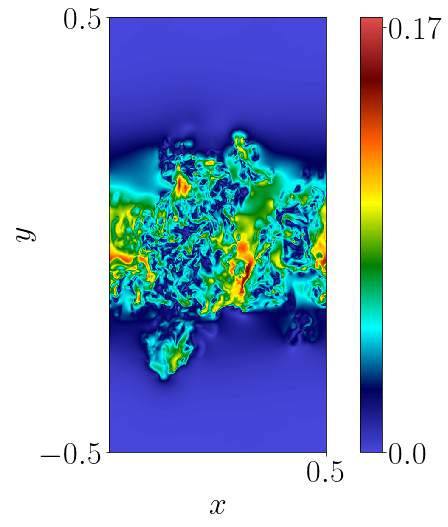}
        \subcaption[]{$A=0.5$}
    \end{subfigure}
    \begin{subfigure}[]{0.35\textwidth}
        \includegraphics[width=\textwidth]{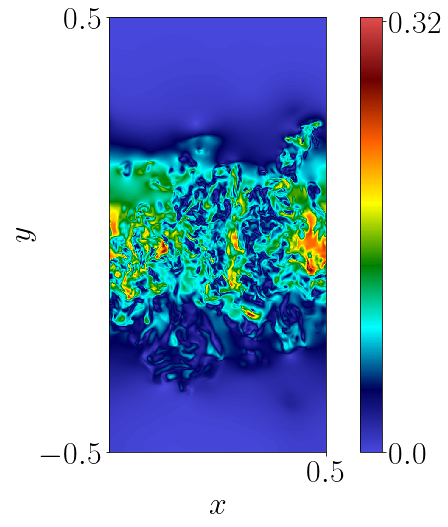}
        \subcaption[]{$A=0.8$}
    \end{subfigure}
    \caption{Contours of Mach number for each $A$ case.}
    \label{fig:contours_mach}
\end{figure}

For statistical convergence, nine realizations are run for each $A$ case.
Unique realizations are achieved by varying seeds for the random number generator.
Example contours of density and mach number from one realization of each Atwood number case are shown in Figures \ref{fig:contours_rho} and \ref{fig:contours_mach}, respectively.

\begin{figure}
    \centering
    \begin{subfigure}[]{0.3\textwidth}
        \includegraphics[height=12em]{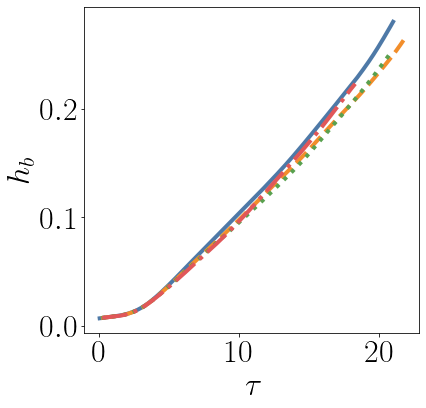}
        \subcaption[]{}
        \label{subfig:hb}
    \end{subfigure}
    \begin{subfigure}[]{0.3\textwidth}
        \includegraphics[height=12em]{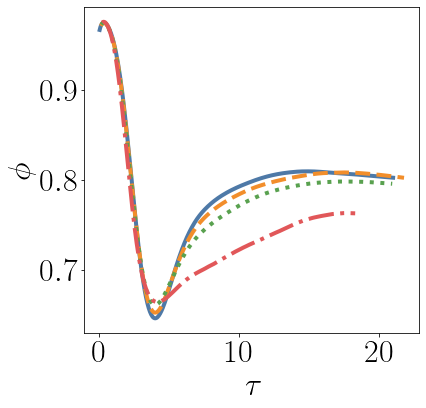}
        \subcaption[]{}
        \label{subfig:phi}
    \end{subfigure}
    \begin{subfigure}[]{0.3\textwidth}
        \includegraphics[height=12em]{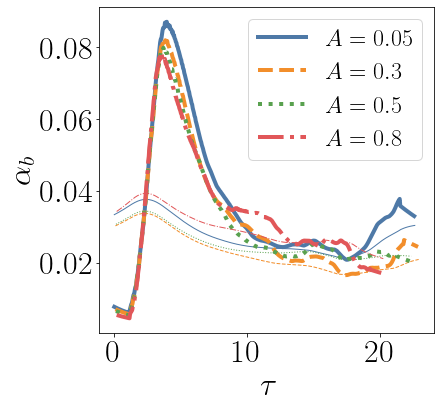}
        \subcaption[]{}
        \label{subfig:alpha}
    \end{subfigure}
    \caption{Self-similarity metrics for each $A$ case\edit{: (a) bubble height, (b) mixedness, (c) bubble growth parameter.
    In (c), thick lines are from Equation \ref{eq:alpha_b}, and thin lines are from Equation \ref{eq:alpha_b_est}.}}
    \label{fig:selfsim_metrics}
\end{figure}

Figure \ref{fig:selfsim_metrics} shows self-similar metrics for the RT simulations, averaged over all realizations for each $A$ case.
\edit{
	$\alpha_b$ computed using the definitions from Equations \ref{eq:alpha_b} and \ref{eq:alpha_b_est} for comparison.
	The $\alpha_b$ computed from Equation \ref{eq:alpha_b} it is not perfectly flat, but this is likely due to statistical error in $h$ and its time derivative.
	On the other hand, $\alpha_b$ computed from Equation \ref{eq:alpha_b_est} is smoother.
	The two definitions do not match at early times when the flow is not self-similar but converge to similar values at late time.}

The $A=0.05$ case appears to be safely in the self-similar regime, as its $\phi$ seems converged to approximately $0.8$.
Its $\alpha_b$ is also somewhat converged to approximately $0.03$, which is within the range reported in the literature \citep{cabotcook2006, livescu2010}.
The $A=0.5$ case also appears to be in the self-similar regime, having converged to similar values of $\alpha_b$ and $\phi$ as the $A=0.05$ case, but the former does not appear to be as far into the self-similar state as the latter.
The $A=0.8$ case gives $\alpha_b$ and $\phi$ that are only beginning to converge, indicating that this case is just barely in the self-similar regime.
Nevertheless, this case gives an $\alpha_b$ close to the other cases and can be used to make quadratic fits for $h_b$ needed for self-similar analysis.

\begin{figure}
	\centering
	\begin{subfigure}[]{0.3\textwidth}
		\includegraphics[height=12em]{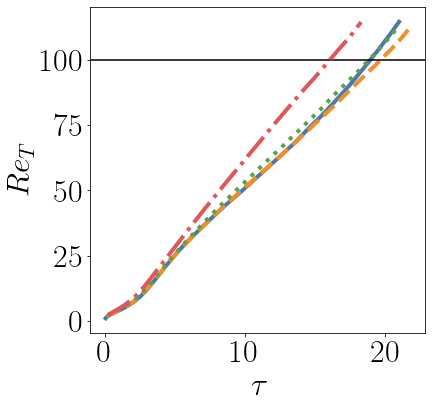}
		\subcaption[]{}
	\end{subfigure}
	\begin{subfigure}[]{0.3\textwidth}
		\includegraphics[height=12em]{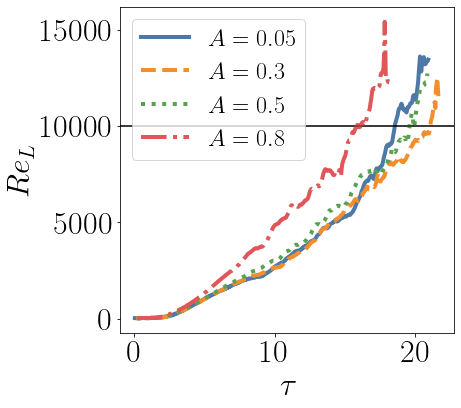}
		\subcaption[]{}
	\end{subfigure}
	\caption{Reynolds numbers over time for each of the $A$ cases.
		The horizontal lines indicate critical values for each Reynolds number.}
	\label{fig:Revst}
\end{figure}

\edit{
The plots of $Re$ in Figure \ref{fig:Revst} show that the flows in all the $A$ cases studied here develop past the critical Reynolds numbers for the mixing transition from \citet{dimotakis2000}.
These plots are given in nondimensional time $\tau=t/\tau_0$, where $\tau_0=\sqrt{h_0/Ag}$, and $h_0$ is the dominant length scale determined by the peak of the initial perturbation spectrum.
The critical $Re$ are reached at approximately $\tau=19$, $\tau=20$, $\tau=19$, and $\tau=16$ for $A=0.05$, $A=0.3$, $A=0.5$, and $A=0.8$, respectively.
Based on this, the RT instabilities simulated here pass the mixing transition.
}

\begin{figure}[t]
	\centering
	\begin{subfigure}[]{0.3\textwidth}
		\includegraphics[width=\textwidth]{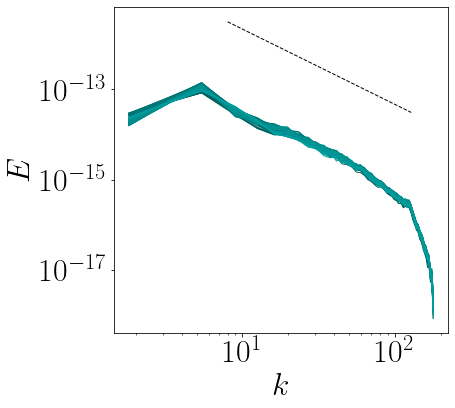}
		\subcaption[]{$A=0.05$}
	\end{subfigure}
	\begin{subfigure}[]{0.3\textwidth}
		\includegraphics[width=\textwidth]{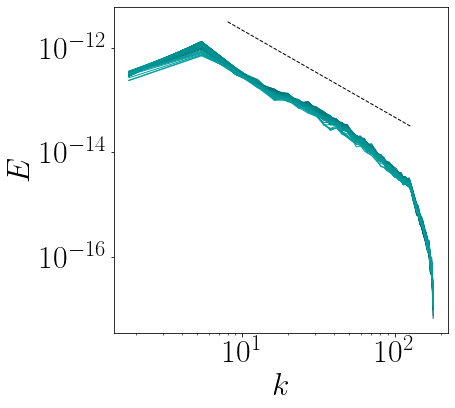}
		\subcaption[]{$A=0.5$}
	\end{subfigure}
	\begin{subfigure}[]{0.3\textwidth}
		\includegraphics[width=\textwidth]{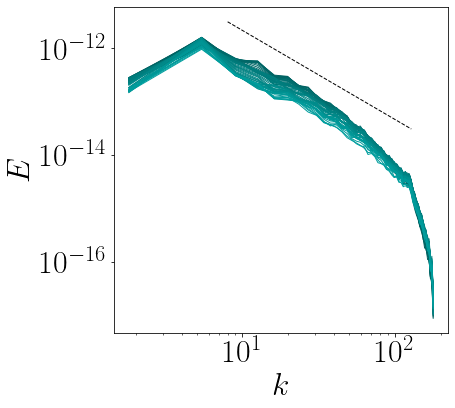}
		\subcaption[]{$A=0.8$}
	\end{subfigure}
	\caption{Energy spectra at different Atwood numbers at the last timesteps of the simulations.
		Different lines are the spectra at different $y$ within the mixing layer.
		Lighter cyan lines are at higher $y$, and darker lines are at lower $y$.
		The dashed black lines have $-5/3$ slopes.}
	\label{fig:energy_spectra}
\end{figure}

\edit{
	We also examine the energy spectra of the different Atwood cases to assess the laminar-turbulent transition.
	To compute the energy spectra, we use the discrete Fourier transform of a quantity $q$ defined as follows:
	\begin{align}
		\widehat{q}(k_x,k_z,y) = \Sigma_{x=0}^{N_x-1}\Sigma_{z=0}^{N_z-1}q(x,y,z)e^{-2\pi i\bigpar{\frac{k_x x}{N_x} + \frac{k_z z}{N_z}}}.
	\end{align}
	The wavenumbers are defined as 
	\begin{align}
		k_x &= \frac{n_x}{L_x},
		\quad k_z = \frac{n_z}{L_z},\\
		n_x &\in \bigbra{-\frac{N_x}{2},\hdots,\frac{N_x}{2}-1},
		\quad n_z \in \bigbra{-\frac{N_z}{2},\hdots,\frac{N_z}{2}-1}.
	\end{align}
	$N_x$ and $N_z$ are the number of mesh points in $x$ and $z$, respectively.
	The energy spectrum is defined as
	\begin{align}
		E_{2D}(k_x,k_z,y)&= \frac{1}{2}\widehat{u''_i}\widehat{u''_i}^* .
	\end{align}
	where $\widehat{u''_i}^*$ is the conjugate of $\widehat{u''_i}$.
	We define the radial wavenumber $k=\sqrt{k_x^2+k_z^2}$ to obtain $E(k,y)$, which is the annular sum of $E_{2D}(k_x,k_z,y)$.
}

\edit{
	The energy spectra within the mixing layer (for $y$ between approximately $0.4$ and $0.6$ cm) are plotted in Figure \ref{fig:energy_spectra}.
	The spectra are taken at the last timestep and for one realization.
	We find that by the time they are stopped, our simulations are fully-turbulent, demonstrated by the existence of an inertial range in their energy spectra, presented in Figure \ref{fig:energy_spectra} in the Appendix.
	This is despite our $Re_L$ from all $A$ cases not attaining the critical Reynolds number for turbulence, $1.6\times 10^5$, identified by \citet{zhou2007}.
}

\begin{figure}[t]
	\centering
	\begin{subfigure}[]{0.3\textwidth}
		\includegraphics[width=\textwidth]{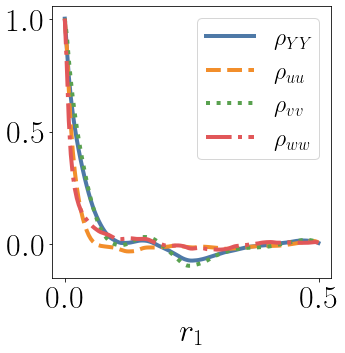}
		\subcaption[]{$A=0.05$}
	\end{subfigure}
	\begin{subfigure}[]{0.3\textwidth}
		\includegraphics[width=\textwidth]{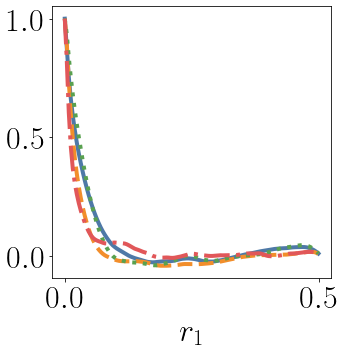}
		\subcaption[]{$A=0.5$}
	\end{subfigure}
	\begin{subfigure}[]{0.3\textwidth}
		\includegraphics[width=\textwidth]{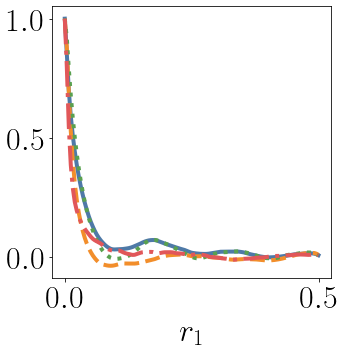}
		\subcaption[]{$A=0.8$}
	\end{subfigure}
	\caption{Correlation curves over $r_1$ at different Atwood numbers at the last timesteps of the simulations.
		Curves are averaged over $z$ and $y$ within the mixing layer.}
	\label{fig:Rii_x}
	\vskip 1em
	\centering
	\begin{subfigure}[]{0.3\textwidth}
		\includegraphics[width=\textwidth]{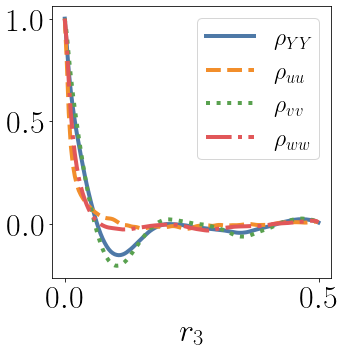}
		\subcaption[]{$A=0.05$}
	\end{subfigure}
	\begin{subfigure}[]{0.3\textwidth}
		\includegraphics[width=\textwidth]{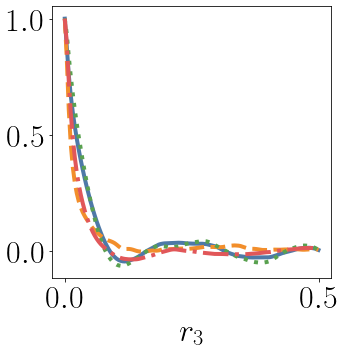}
		\subcaption[]{$A=0.5$}
	\end{subfigure}
	\begin{subfigure}[]{0.3\textwidth}
		\includegraphics[width=\textwidth]{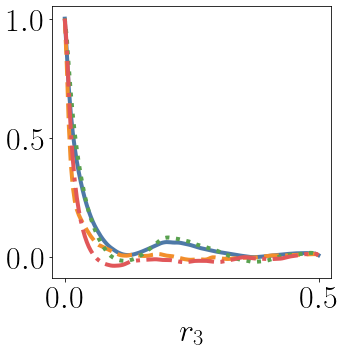}
		\subcaption[]{$A=0.8$}
	\end{subfigure}
	\caption{Correlation curves over $r_3$ at different Atwood numbers at the last timesteps of the simulations.
		Curves are averaged over $x$ and $y$ within the mixing layer.}
	\label{fig:Rii_z}
\end{figure}

\edit{
	Additionally, we examine the normalized autocorrelation curves over $x$ and $z$, which are defined by \citet{pope2001turbulent} for a quantity $q$ as
	\begin{align}
		\rho_{qq}(r_1) = \frac{\langle q(x,y,z,t)q(x+r_1,y,z,t)\rangle}{\langle q(x,y,z,t)^2\rangle},\\
		\rho_{qq}(r_3) = \frac{\langle q(x,y,z,t)q(x,y,z+r_3,t)\rangle}{\langle q(x,y,z,t)^2\rangle}.
	\end{align}
	In this case, $\langle * \rangle$ denotes averaging over the remaining homogeneous direction and $y$ within the mixing layer.
	Figures \ref{fig:Rii_x} and \ref{fig:Rii_z} show the autocorrelation curves for $Y_H''$ (denoted by $\rho_{YY}$) and $u_i''$ (denoted by $\rho_{uu}$, $\rho_{vv}$, and $\rho_{ww}$)over $x$ and $z$, respectively, at the end of one realization for each case.
	We observe the curves decay to zero, indicating the simulations stop before the flow encounters lateral confinement effects.
	\editt{
	This is supported by the energy spectra in Figure \ref{fig:energy_spectra}, which demonstrate no evidence of turbulent kinetic energy saturation at low wavenumbers.
	}
	We include the autocorrelations in $y$ and the associated integral length scales in Appendix \ref{appendix:autocorr_y}.
}

\section{Analysis of nonlocality}
\label{sec:nonlocality}

\subsection{Modeling the mean scalar transport operator}
\label{subsec:mstoper}
To obtain the mean scalar transport equation, the Reynolds ($\ens{q}$) and Favre ($\widetilde{q}$) averages of quantity $q$ are defined:
\begin{align}
	\ens{q} &= \frac{1}{N}\sum_i^N q_i,\\
	\widetilde{q} &= \frac{\ens{\rho q}}{\rhom},
\end{align}
where $N$ is the number of ensembles.
In the case where the flow is homogeneous (in space and/or time), the homogeneous directions may be included in these ensembles.
For the RT mixing problem studied here, the homogeneous directions are $x$ and $z$, so averages are performed over $x$, $z$, and realizations.
Fluctuations from the Reynolds and Favre means are denoted as $q'$ and $q^{''}$, respectively, so
\begin{align}
	q = \ens{q}+q' = \widetilde{q}+q^{''}.
\end{align}

Substituting the Favre decomposition for velocity and mass fraction into Equation \ref{eq:scalar_transport} and taking its Reynolds average results in the mean scalar transort equation for compressible flow:
\begin{align}
	\diffp{\rhom\YHm}{t}=-\diffp{}{y}\bigpar{\rhom\widetilde{v}\YHm + \ens{\rho v^{''}Y_H^{''}} - \ens{\rho D_H \diffp{Y_H}{y}}}.
\end{align}
The last term on the right hand side is negligible when $Pe$ is large.
The turbulent species flux $F=\tsf$ is unclosed and needs to be modeled.
This flux can be exactly expressed as
\begin{align}
	F\bigpar{y,t} = \rhom\int D\bigpar{y,y',t,t'}\left.\diffp{\YHm}{y}\right|_{y',t'}dy'dt',
\end{align}
where $D$ is the eddy diffusivity kernel.
This equation is an extension of the formulation for incompressible flow described in \citet{kraichnan1987}, \citet{hamba2005}, and \citet{manipark2021} .
It is a nonlocal formulation, in that it expresses $F$  based on mean scalar gradients not only at the points in space and time ($y$ and $t$) at which $F$ is measured, but also all other points in space and time ($y'$ and $t'$).
\edit{Numerical tests have shown that this formulation satisfies causality without imposing it \citep{liu2024}.}

An exact model for $F$ requires full characterization of the eddy diffusivity kernel. This has been done for simpler flows in the works of \citet{hamba1995}, \citet{hamba2005}, and \citet{park2024}. 
However, computation of the kernel is generally computationally expensive, since it requires simulations on the order of the number of points in macroscopic space.
On top of this, chaotic flows, like RT mixing, require many realizations for statistical convergence.

The eddy diffusivity kernel can instead be approximated by its moments.
This can be done by employing a Taylor series expansion of the mean scalar gradient about $y$ and $t$, which results in the Kramers-Moyal-like expansion:

\begin{align}
	F\bigpar{y,t} &= 
	\rhom D^{00}(y,t) \diffp{\YHm}{y}
	+ \rhom D^{10}(y,t) \diffp{\YHm}{{y^2}}
	+ \rhom D^{01}(y,t) \diffp{\YHm}{ty}\nonumber\\
	&+ \rhom D^{20}(y,t) \diffp{\YHm}{{y^3}}
	+ \rhom D^{11}(y,t) \diffp{\YHm}{{ty^2}}
	+ \rhom D^{02}(y,t) \diffp{\YHm}{{t^2y}}
	+ \dots
	\label{eq:KME}
\end{align}
where $D^{mn}$ are the eddy diffusivity moments.
The first index $m$ indicates space and the second index $n$ is time.
The moments are defined as
\begin{align}
	D^{mn}(y,t) &= \int \int \frac{(y'-y)^m(t'-t)^n}{m!n!}D(y,y',t,t')dy' dt'.
\end{align}
These moments are more computationally feasible to compute than the full kernel.
To compute the moments, which will be described shortly, one equation per moment needs to be added to the suite of equations being solved in a simulation.
Though the number of operations increases as more moments are computed, only one simulation needs to be run to compute all moments.
Statistically converged moments require multiple simulations; in this work, it is found that $\mathcal{O}(10)$ simulations are needed for statistical convergence sufficient for analysis, which is much lower than what is needed to compute the full kernel.

While the eddy diffusivity moments are locally defined (they are functions of $y$ and $t$ only), higher-order moments contain information about the nonlocality of the full kernel.
The leading-order moment $D^{00}$ is purely local, and truncation to the leading-order term is the gradient-diffusion or Boussinesq approximation.
The goal of this work is to determine the importance of the higher-order terms and, therefore, the nonlocality of the mean scalar transport operator for the RT mixing cases studied here.
In this way, measuring the eddy diffusivity moments is more computationally efficient than computing the full kernel, but it is still an insightful way to assess the nonlocality of the closure operator.

\subsection{Measuring the eddy diffusivity moments using the Macroscopic Forcing Method}
\label{subsec:mfm}

\begin{figure}
	\centering
	\resizebox{\linewidth}{!}{
\begin{tikzpicture}[
	node distance=2em, 
	auto,
	rect/.style={
		rectangle, 
		inner sep=1em
	},
	rrect/.style={
		rect, 
		rounded corners=10
	},
	smarrow/.style={
		arrows = {-Latex[width'=0pt .5, length=10pt, 
			line width=1.5pt]}
	}]
	\tikzset{>=latex}
	\pgfdeclarelayer{back}
	\pgfsetlayers{back,main}
	
	\node[rect] (base) {};
	\node[below left=0em and 1.5em of base.north, rect,fill=lightblue,draw=lightblue,anchor=north] (donor) {Donor};
	\node[below right=.75em and 1.5em of donor,rect,draw=seagreen,fill=white,line width=1.5pt, minimum width=12em] (n0) {$\frac{D \cmom^{00}}{D t}=\mathcal{L}^{00}(\cmom^{00})+s^{00}$};
	\draw [arrows = {-Stealth[inset=0pt, angle=90:2em]},lightblue, line width=1em] (donor) |- (n0)[midway];
	\node[below=2em of n0,rect,draw=seagreen,fill=white,line width=1.5pt, minimum width=12em] (n1) {$\frac{D \cmom^{10}}{D t}=\mathcal{L}^{10}(\cmom^{10})+s^{10}$};
	\draw [arrows = {-Stealth[inset=0pt, angle=90:2em]},lightblue, line width=1em] (donor) |- (n1)[midway];
	\node[below=2em of n1,rect,draw=seagreen,fill=white,line width=1.5pt, minimum width=12em] (n2) {$\frac{D \cmom^{20}}{D t}=\mathcal{L}^{20}(\cmom^{20})+s^{20}$};
	\draw [arrows = {-Stealth[inset=0pt, angle=90:2em]},lightblue, line width=1em] (donor) |- (n2)[midway];
	
	\node[anchor=north east, at=(donor.north west),xshift=-3em, rect, draw=lightblue, align=center, line width=1.5pt] (NS) {
		Navier--Stokes Equations
		\\  \\
		$
		\begin{aligned}
			\frac{\partial \rho Y_\alpha}{\partial t}=&-\frac{\partial }{\partial x_j} 		\bigpar{ \rho u_j Y_\alpha - \rho D_\alpha\frac{\partial Y_\alpha}{\partial 		x_j} }\\
			\frac{\partial \rho u_i}{\partial t}=&-\frac{\partial }{\partial x_j} \bigpar{ 		\rho u_i u_j + p\delta_{ij} - \tau_{ij} } + {\rho g_i}\\
			\frac{\partial E}{\partial t}=&-\frac{\partial }{\partial x_j} \bigpar{ 		\bigpar{E+p}u_j - \tau_{ij}u_i + q_i } + \rho g_i u_i\\
			\rho =& \rho Y_H + \rho Y_L
		\end{aligned}
		$
	};
	\draw[lightblue, line width=1.5pt] (donor.west) -- ++(-3em,0);
	
	\begin{pgfonlayer}{back}
		\node[] at (donor.north -| n0.north) (aligner) {};
	\end{pgfonlayer};
	\node[inner sep=0,above=1em of n0.north west,anchor=north west] (receivers) {\color{seagreen}Receiver};
	\node[inner sep=0,above=1em of n1.north west,anchor=north west] (receivers) {\color{seagreen}Receiver};
	\node[inner sep=0,above=1em of n2.north west,anchor=north west] (receivers) {\color{seagreen}Receiver};
	
	\node[left=4.5em of n0.west] (s00) {$\frac{\partial \langle c\rangle}{\partial x}=1$};
	\draw[smarrow] (s00) -> (n0);
	\node[left=4.5em of n1.west] (s10) {$\frac{\partial \langle c\rangle}{\partial x}=x$};
	\draw[smarrow] (s10) -> (n1);
	\node[left=4.5em of n2.west] (s20) {$\frac{\partial \langle c\rangle}{\partial x}=\frac{x^2}{2}$};
	\draw[smarrow] (s20) -> (n2);

	\begin{pgfonlayer}{back}
		\node[
		rrect,
		minimum height=20em,
		minimum width=12em,
		above right=1em and 21.5em of base.north,
		anchor=north,
		draw=black!15,
		line width=1.5pt,
		dashed
		] (post_back) {};
	\end{pgfonlayer};
	\node[inner sep=0,below left=1em and 0em of post_back.north,anchor=north,align=center] (post) {\\ \textcolor{black!50}{  \textit{Postprocessing}}};
	
	\begin{pgfonlayer}{back}
		\node[
		rrect,
		minimum height=20em,
		minimum width=19.5em,
		left=16em of post_back.north,
		anchor=north,
		draw=black!15,
		line width=1.5pt,
		dashed
		] (dns_back) {};
	\end{pgfonlayer};
	\node[inner sep=0,below left=1em and 0em of dns_back.north,anchor=north,align=center] (dns) {\textcolor{black!50}{\textit{Numerical}}\\[-0.5em] \textcolor{black!50}{\textit{Simulation}}};
	
	\node[rect, draw=orange, line width=1.5pt, right=2.5em of n0.east] (F00) {$D^{00}=\frac{\ens{-\rho v''\cmom^{00}}}{\rhom}$};
	\draw[smarrow] (n0) -> (F00);
	\node[rect, draw=orange, line width=1.5pt, right=2.5em of n1.east] (F10) {$D^{10}=\frac{\ens{-\rho v''\cmom^{10}}}{\rhom}$};
	\draw[smarrow] (n1) -> (F10);
	\node[rect, draw=orange, line width=1.5pt, right=2.5em of n2.east] (F20) {$D^{20}=\frac{\ens{-\rho v''\cmom^{20}}}{\rhom}$};
	\draw[smarrow] (n2) -> (F20);
	
\end{tikzpicture}}
	\caption{MFM pipeline illustrating measurement of some spatial eddy diffusivity moments for the mean scalar transport problem. 
	$\mathcal{L}^{mn}$ are the right-hand side operators corresponding to the $\cmom^{mn}$ in Equations \ref{eq:decomp_start}-\ref{eq:decomp_end}, not including the macroscopic forcings $s^{mn}$.}
	\label{fig:MFM}
\end{figure}
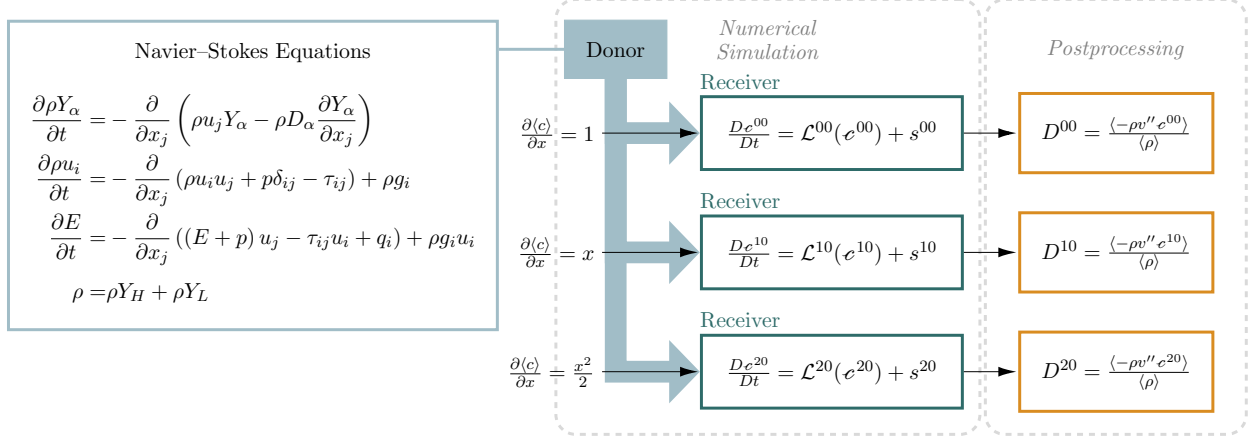

\begin{table}[]
	\centering
	{\renewcommand{\arraystretch}{1.15}%
		\begin{tabular}{c l}
			Moment &  $\diffp{\Ycm}{y}$\\
			\hline
			$D^{00}$ & $1$\\
			$D^{01}$ & $t$\\
			$D^{10}$ & $y-\frac{1}{2}$\\
			$D^{02}$ & $\frac{1}{2}t^2$\\
			$D^{11}$ & $(y-\frac{1}{2})t$\\
			$D^{20}$ & $\frac{1}{2}(y-\frac{1}{2})^2$\\ 
	\end{tabular}}
	\caption{Mean mass fraction gradients forced for each eddy diffusivity moment $D^{mn}$.}
	\label{tab:moment_forcings}
\end{table}

The MFM pipeline is conceptually illustrated in Figure \ref{fig:MFM}.
The method involves two sets of equations, called the \textit{donor} and the \textit{receiver} equations, which are solved simultaneously in a simulation.
The donor equations represent the full set of governing equations (in this case, Equations \ref{eq:scalar_transport}-\ref{eq:NS_end}) and provides quantities necessary for solution of the receiver equations, which involve additional forcing terms $s^{mn}$ corresponding to each moment $D^{mn}$.
For example, to determine closures for scalar transport, 
\edit{
	forcings are applied to the dilute scalar transport equation, Equation \ref{eq:ste_dilute}, resulting in the following receiver equations:
\begin{align}
	\frac{\partial \rho Y_c^{mn}}{\partial t}&=-\frac{\partial }{\partial x_j} \bigpar{ \rho u_j Y_c^{mn} - \rho D_c\frac{\partial Y_c^{mn}}{\partial x_j} }+s^{mn},\label{eq:ste_forced}
\end{align}
where $Y_c^{mn}$ is the dilute scalar field response to the forcing $s^{mn}$ used to obtain $D^{mn}$.
}
The solution of these receiver equations require velocity and density fields, which are obtained from the donor equations.
A key component of this method is that the forcings are macroscopic, in that they have the property $\ens{s^{mn}}=s^{mn}$, so the forcings do not interfere with the underlying turbulent flow.

In the present work, a new decomposition approach to the macroscopic forcing method is utilized to efficiently obtain the moments. 
Full details on this method can be found in the companion work by \cite{lavacot2025}. 
Under this new method, fluctuations of the mass fraction field are simulated, and the forcings are semi-analytically applied to achieve certain mean scalar gradients that allow for the probing of each eddy diffusivity moment.
Table \ref{tab:moment_forcings} lists the mean scalar gradients chosen to obtain each moment.

To formulate the forced equations, $Y_c''$ is first written as a Kramers-Moyal expansion analagous to Equation \ref{eq:KME}:
\begin{align}
	Y_c'' &= 
	\cpp{00} \diffp{\Ycm}{y}
	+ \cpp{10} \diffp{\Ycm}{{y^2}}
	+ \cpp{01} \diffp{\Ycm}{ty}
	+ \cpp{20} \diffp{\Ycm}{{y^3}}
	+ \cpp{11} \diffp{\Ycm}{{ty^2}}
	+ \cpp{02} \diffp{\Ycm}{{t^2y}}
	+ \dots
	\label{eq:KME_cpp}
\end{align}
Substituting this expansion and the forced mean mass fractions derived from Table \ref{tab:moment_forcings} into Equation \ref{eq:scalar_transport} gives the following equations for each $\cpp{mn}$:
\begin{align}
	\DDt{\rho\cpp{00}} &= \diffu{\cpp{00}} - \rho v + \diffp{}{{y}}\rho D_H + s^{00},  \label{eq:decomp_start}\\
	\DDt{\rho\cpp{10}} &= \diffu{\cpp{10}} - \rho v\cpp{00} + \rho D_H  + \rho D_H \diffp{}{{y}}\cpp{00}  + \diffp{}{{y}}\rho D_H \cpp{00} + s^{10},  \\
	\DDt{\rho\cpp{01}} &= \diffu{\cpp{01}} - \rho\cpp{00} - \rho y + s^{01}, \\
	\DDt{\rho\cpp{20}} &= \diffu{\cpp{20}} + \rho D_H \cpp{00} + \rho D_H \diffp{}{{y}}\cpp{10} + \diffp{}{{y}}\rho D_H \cpp{10} + s^{20},\\
	\DDt{\rho\cpp{11}} &= \diffu{\cpp{11}} - \rho\cpp{10} - \rho v \cpp{01} + \rho D_H \diffp{}{{y}}\cpp{01} + \diffp{}{{y}}\rho D_H \cpp{01} + \rho\bigpar{\frac{1}{2}y^2-\frac{1}{8}} + s^{11},\\
	\DDt{\rho\cpp{02}} &= \diffu{\cpp{02}} - \rho \cpp{01} + s^{02},\label{eq:decomp_end}
\end{align}
where each forcing $s^{mn}$ enforces the $x$-$z$ mean of $\cpp{mn}$ to be zero, and
\edit{
	$\cpp{mn}$ are initially zero.
}
To obtain the eddy diffusivity moments, the turbulent species flux based on the $Y_c''$ from each equation is computed in postprocessing:
\begin{align}
	\ens{-\rho v''\cpp{mn}}=\rhom D^{mn}.
\end{align}

In the numerical simulation, solutions to the donors (Equations \ref{eq:scalar_transport}-\ref{eq:NS_end}) are given to these receiver equations, which are solved alongside the donors.
Thus, if the cost to solve the suite of donor equations is $N$, the cost of MFM for the number of eddy diffusivity moments examined in this work is approximately $2N$.
Of course, this cost increases as more moments are measured, but it has been found that not many moments are required to characterize the nonlocality of the eddy diffusivity kernel \citep{lavacot2024, liu2023}, making the MFM measurement of moments relatively efficient and useful.
\subsection{Self-similar scaling}
\label{subsec:selfsimnorm}
\begin{figure}
	\centering
	\begin{subfigure}[]{0.24\textwidth}
		\includegraphics[width=\textwidth]{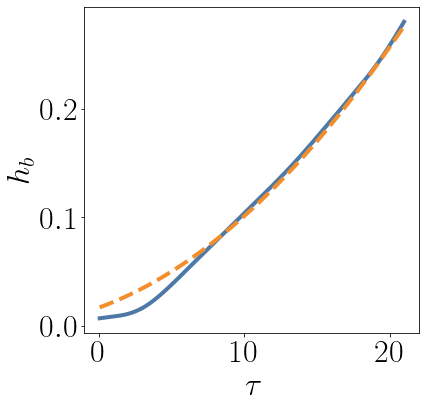}
		\subcaption[]{$A=0.05$}
		\label{subfig:hb}
	\end{subfigure}
	\begin{subfigure}[]{0.24\textwidth}
		\includegraphics[width=\textwidth]{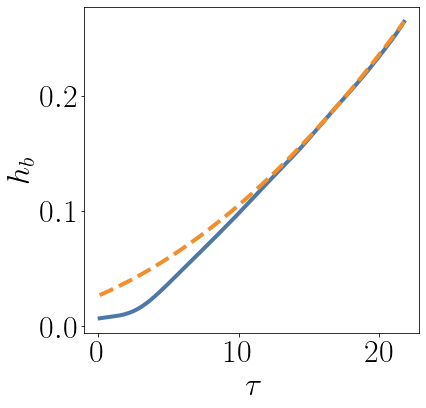}
		\subcaption[]{$A=0.3$}
		\label{subfig:alpha}
	\end{subfigure}
	\begin{subfigure}[]{0.24\textwidth}
		\includegraphics[width=\textwidth]{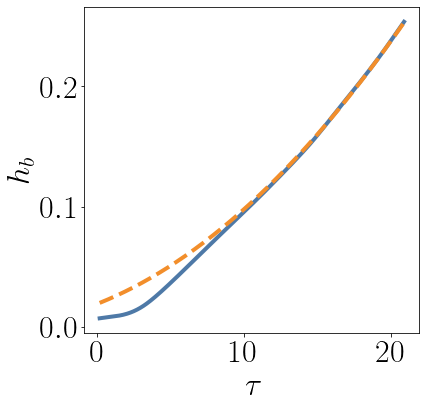}
		\subcaption[]{$A=0.5$}
		\label{subfig:alpha}
	\end{subfigure}
	\begin{subfigure}[]{0.24\textwidth}
		\includegraphics[width=\textwidth]{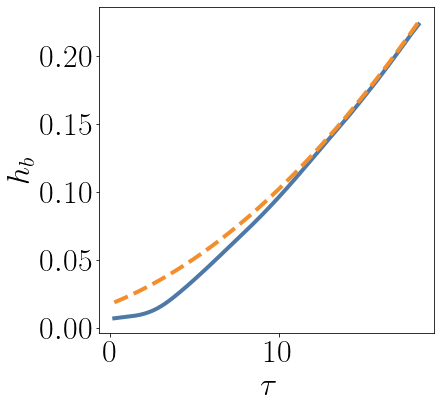}
		\subcaption[]{$A=0.8$}
		\label{subfig:phi}
	\end{subfigure}
	\caption{$h_b$ from simulations (solid blue) and fitted $h_b$ (dashed orange).}
	\label{fig:h_fit}
\end{figure}

\citet{lavacot2024} utilize a self-similar normalization for the analysis of $F$ and eddy diffusivity moments.
This analysis is applied directly to the variable density RT mixing studied here and extended to the higher-order moments $D^{11}$ and $D^{02}$, which have not been studied previously:
\begin{align}
	\widehat{F} =& \frac{F}{\alpha^*Ag (t-t_0^*)},\label{eq:F_selfsim}\\
	\widehat{D^{00}} =& \frac{D^{00}}{(\alpha^*Ag)^2 (t-t_0^*)^3},\label{eq:selfsim1}\\
	\widehat{D^{01}} =& \frac{D^{01}}{(\alpha^*Ag)^2 (t-t_0^*)^4},\\
	\widehat{D^{10}} =& \frac{D^{10}}{(\alpha^*Ag)^3 (t-t_0^*)^5},\\
	\widehat{D^{02}} =& \frac{D^{02}}{(\alpha^*Ag)^2 (t-t_0^*)^5},\\
	\widehat{D^{11}} =& \frac{D^{11}}{(\alpha^*Ag)^3 (t-t_0^*)^6},\\
	\widehat{D^{20}} =& \frac{D^{20}}{(\alpha^*Ag)^4 (t-t_0^*)^7},\label{eq:selfsim2}
\end{align}
where $\alpha^*$ is the growth parameter defined in Equation \ref{eq:alpha_b_est}, and $t_0^*$ is a fitted time origin based on the measured bubble height.
Figure \ref{fig:h_fit} shows the bubble height measured from the simulations and the determined fits, which aim to match $h_b$ in late time (during self-similar growth).

In addition, since finite $A$ are considered in this work, a self-similar scaling for the peak mean velocity $V_0(t)$ is also used:
\begin{align}
	\widehat{V_0} = \frac{V_0}{Ag(t-t_0^*)}.
\end{align}

Due to this nonzero $\vm$ at finite $A$, there is asymmetry in the RT mixing layer, and its midplane shifts in the negative $y$ direction.
This shift is incorporated into the self-similar variable:
\begin{align}
	\eta=\frac{\bigpar{y-\frac{1}{2}}-V_0 (t-t_0^*)}{h_b}=
	\frac{\bigpar{y-\frac{1}{2}}}{h_b}-\frac{ \widehat{V_0} }{\alpha^*}.
	\label{eq:eta_shift}
\end{align}
All $\eta$ in the following analyses are defined this way.

\begin{figure}
	\centering
	\begin{subfigure}[]{0.24\textwidth}
		\includegraphics[height=9em]{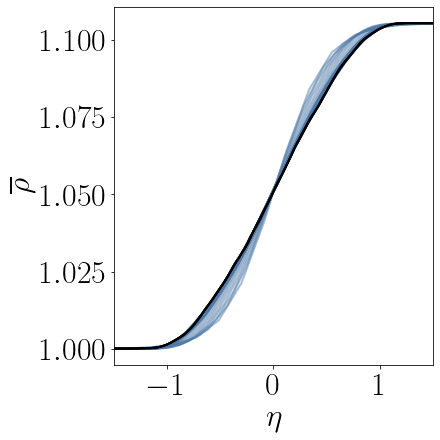}
		\subcaption[]{$A=0.05$}
		\label{subfig:rho_selfsim_A005}
	\end{subfigure}
	\begin{subfigure}[]{0.24\textwidth}
		\includegraphics[height=9em]{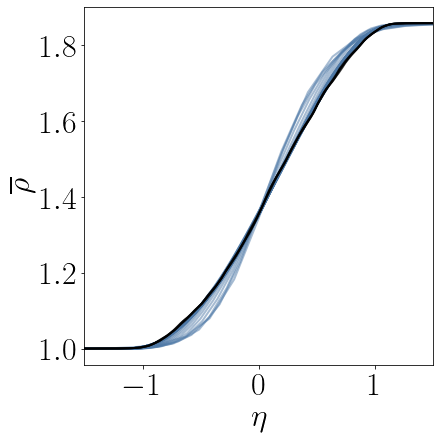}
		\subcaption[]{$A=0.3$}
		\label{subfig:rho_selfsim_A03}
	\end{subfigure}
	\begin{subfigure}[]{0.24\textwidth}
		\includegraphics[height=9em]{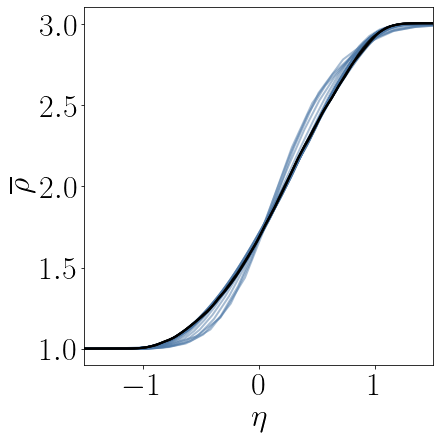}
		\subcaption[]{$A=0.5$}
		\label{subfig:rho_selfsim_A05}
	\end{subfigure}
	\begin{subfigure}[]{0.24\textwidth}
		\includegraphics[height=9em]{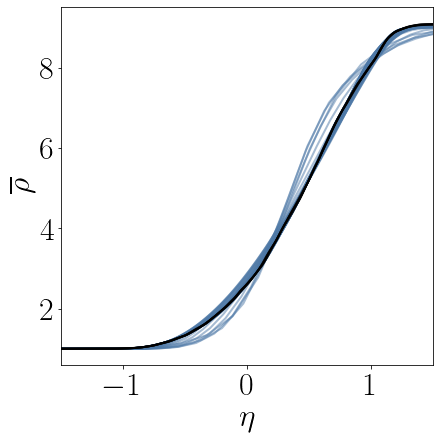}
		\subcaption[]{$A=0.8$}
		\label{subfig:rho_selfsim_A08}
	\end{subfigure}
	\caption{Self-similar collapse of $\rho$.
	Blue lines are before the transition to turbulence, and black lines are after. }
	\label{fig:rho_selfsim}
\end{figure}

\begin{figure}
\centering
\begin{subfigure}[]{0.24\textwidth}
	\includegraphics[height=9em]{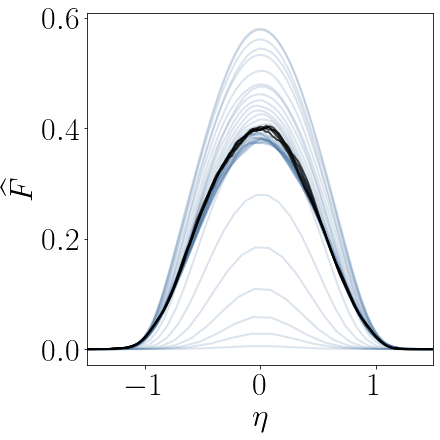}
	\subcaption[]{$A=0.05$}
	\label{subfig:tsf_selfsim_A005}
\end{subfigure}
\begin{subfigure}[]{0.24\textwidth}
	\includegraphics[height=9em]{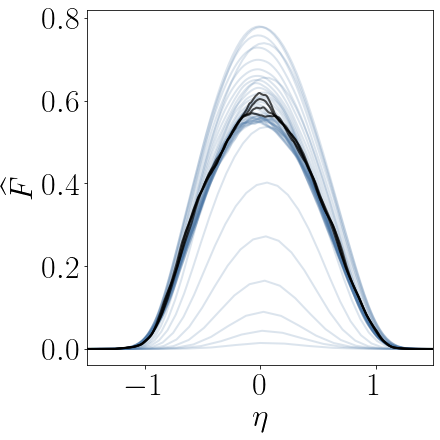}
	\subcaption[]{$A=0.3$}
	\label{subfig:tsf_selfsim_A03}
\end{subfigure}
\begin{subfigure}[]{0.24\textwidth}
	\includegraphics[height=9em]{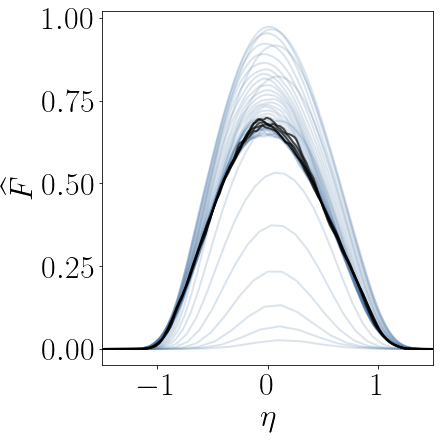}
	\subcaption[]{$A=0.5$}
	\label{subfig:tsf_selfsim_A05}
\end{subfigure}
\begin{subfigure}[]{0.24\textwidth}
	\includegraphics[height=9em]{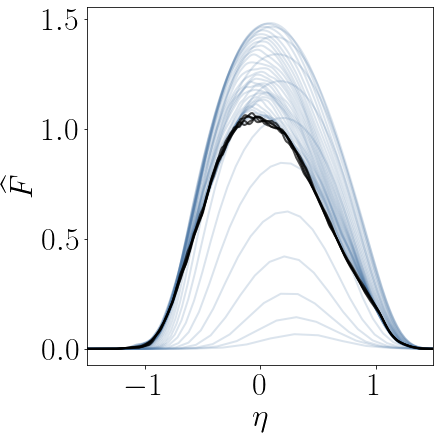}
	\subcaption[]{$A=0.8$}
	\label{subfig:tsf_selfsim_A08}
\end{subfigure}
\caption{Self-similar collapse of $\widehat{F}$.
	Blue lines are before the transition to turbulence, and black lines are after.}
\label{fig:tsf_selfsim}
\end{figure}

Figures \ref{fig:rho_selfsim} and \ref{fig:tsf_selfsim} show the density and the turbulent species flux profiles, respectively, using the $\eta$ defined in \ref{eq:eta_shift}.
The flux is normalized according to \ref{eq:F_selfsim}.
In these self-similar coordinates, the profiles collapse \edit{in late time}; these profiles are represented by the black lines in the figures.
In the analyses to follow, the self-similar profiles from the last timesteps of the simulations are used.

\subsection{Eddy diffusivity moments}
\label{subsec:moments}

\begin{figure}
    \centering
    \resizebox{.75\linewidth}{!}{\input{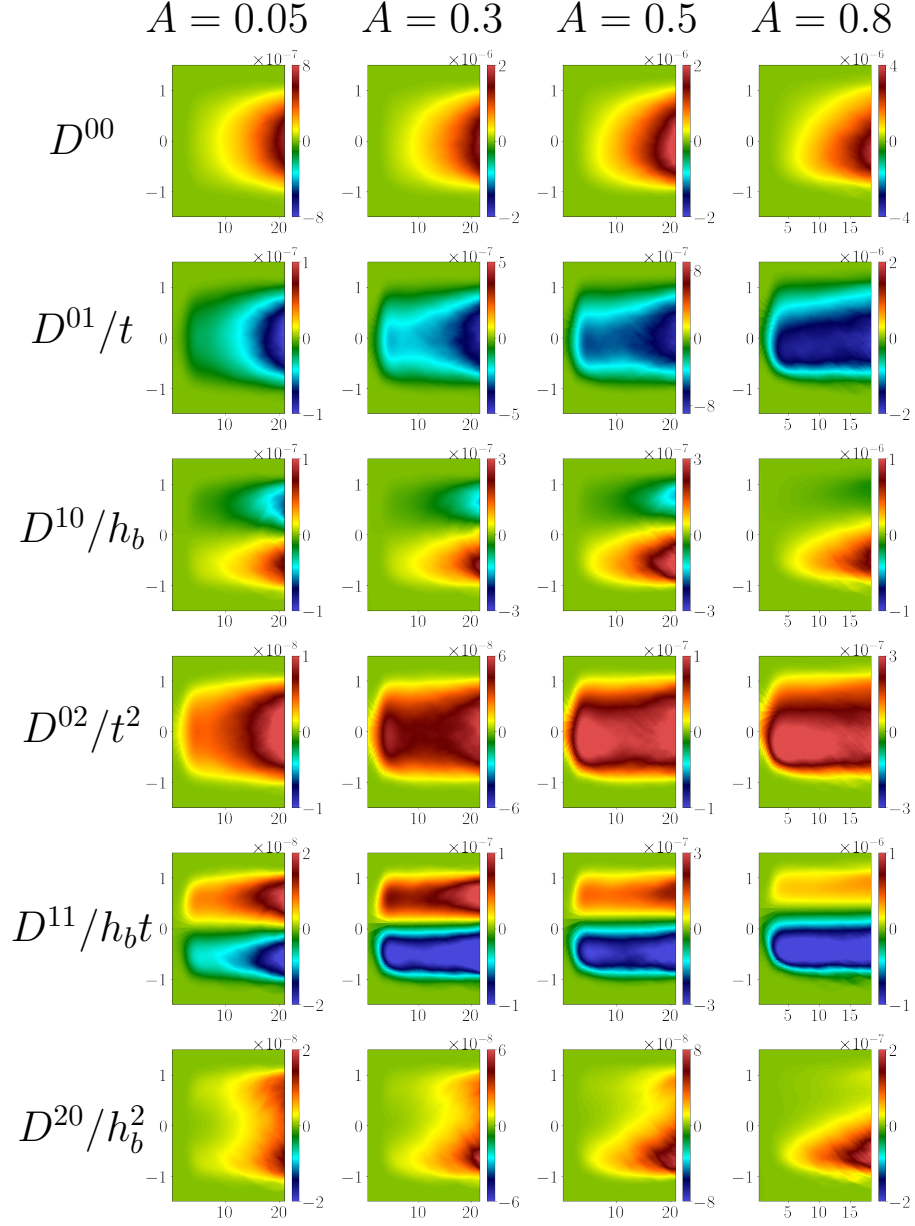}}
    \caption{Eddy diffusivity moments of RT instability at different Atwood numbers.
    The $y$ axis is $\eta$, and the $x$ axis is $\tau$.
    Moments are normalized by appropriate length and timescales so that all dimensions match.}
    \label{fig:moments}
\end{figure}

Figure \ref{fig:moments} shows the eddy diffusivity moments for each Atwood number.
The vertical axis is the self-similar variable $\eta$, and the horizontal axis is the nondimensional time $\tau$.
Each row corresponds to an eddy diffusivity moment, and Atwood number increases across the columns from left to right.
The contours show the values of the eddy diffusivity moments at each $\eta$ and $\tau$, normalized by length and time scales such that units are the same across plots.
First, some expected behavior is observed at the lowest Atwood number:
\begin{enumerate}
    \item $D^{00}$ is the largest in magnitude. It is also symmetric and positive.
    \item $D^{10}$ is antisymmetric. 
    $D^{10}$ is negative above the centerline ($\eta=0$), indicating that at a point $y_0$ above the centerline, mixing depends more on gradients closer to the centerline rather than the edge of the mixing layer.
    That is, the kernel has a centroid that is a negative distance away from $y_0$ above the centerline.
    Similarly, $D^{10}$ is positive below the centerline.
    \item $D^{01}$ is symmetric and always negative. This is expected based on causality.
    \item $D^{11}$ is antisymmetric.
    \item $D^{02}$ and $D^{20}$ are symmetric and always positive, which is characteristic of the moment of inertia of a positive kernel.
\end{enumerate}

As $A$ increases, the moments become asymmetric.
This is expected for $A$ above about $0.1$, since at these higher density differences, the heavy fluid falls deeper than the light fluid rises, moving the mixing layer center line downward.
The asymmetry of RT instability with finite Atwood numbers is well known, and it has also been found that quantities in turbulence budgets (e.g., mass flux and turbulent kinetic energy) are skewed in these regimes \citep{livescu2010}.
Thus, peak magnitudes of moments that are symmetric at low $A$ move further below the domain center line as Atwood increases.
Similarly, below the centerline, the magnitudes of moments that are antisymmetric at low $A$ become larger than the magnitudes above the centerline.

\begin{figure}
    \centering
    \resizebox{.75\linewidth}{!}{\input{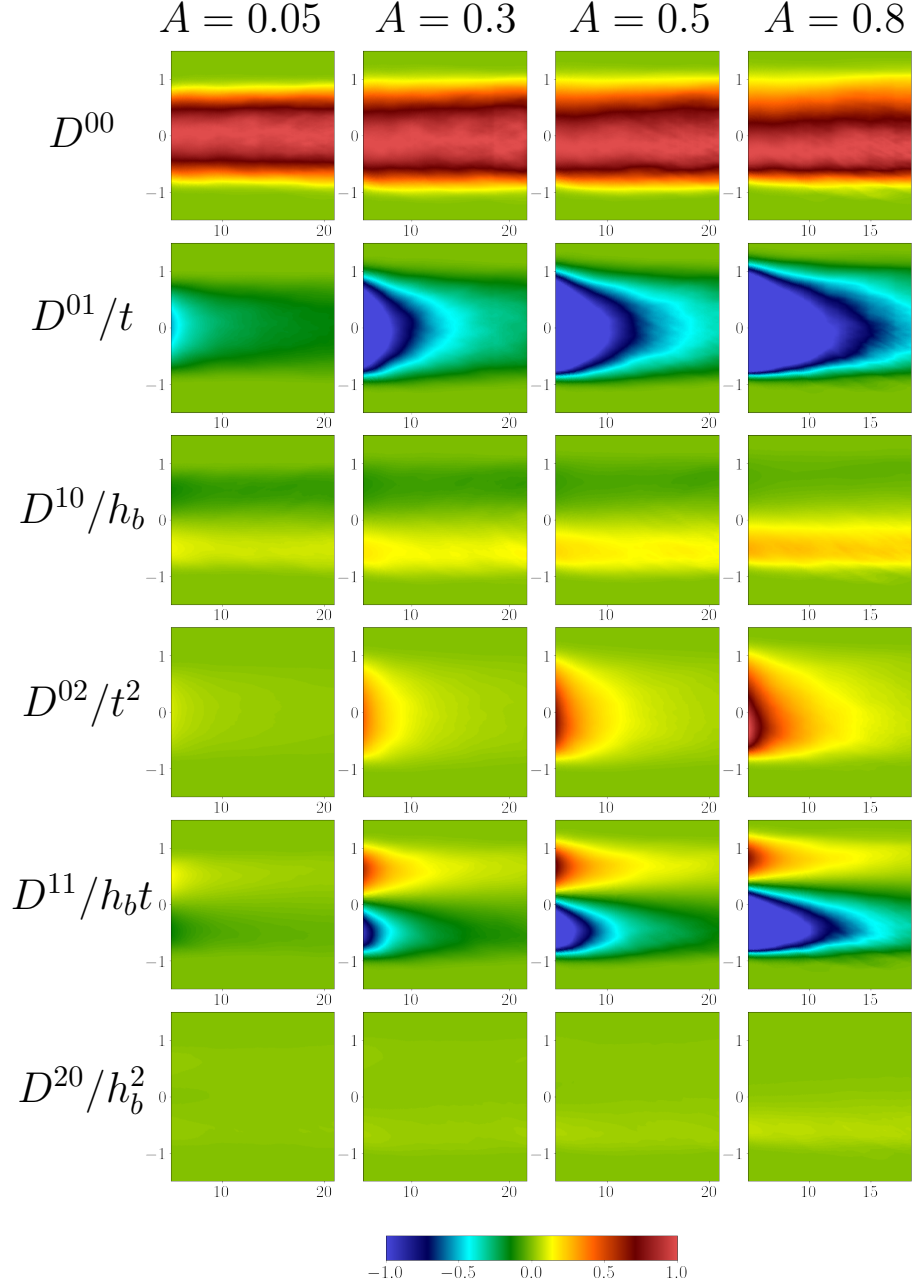}}
    \caption{\edit{
    	Eddy diffusivity moments normalized by the  maximum magnitude of the leading order moment at every time slice for each Atwood number.
    	The  $y$ axis is $\eta$, and the $x$ axis is $\tau$.
    	Data for $\tau>5$ is shown.
	    }
	}
    \label{fig:rel_moments}
\end{figure}

A preliminary assessment of nonlocality can be done by examining the relative magnitudes of the measured moments, which is shown in Figure \ref{fig:rel_moments}.
Maximum relative magnitude values are also provided in Table \ref{tab:rel_moments}.
\edit{
	For all Atwood numbers, the magnitudes of the first order moments are about one magnitude smaller than the leading order moment; the second order moments are about two orders of magnitude smaller than the leading order moment.
	This suggests that these higher-order moments may be significant and may not be excluded from modeling right away.
}
This was also observed by \citet{lavacot2024} for 2D RT instability at $A=0.05$.
\edit{
	More notable are the changes in relative magnitude over the Atwood cases and over time.
	Specifically, the following trends are observed:
	\begin{itemize}
		\item The spatial moments increase in relative magnitude with $A$.
		\item The temporal moments are higher in relative magnitude at early times.
		\item The temporal moments appear to increase in relative magnitude with $A$, but this is difficult to quantify due to the temporal decay of temporal moments.
		\item The length of time over which the temporal moments decay increases with $A$, suggesting that history effects last longer with higher $A$.
	\end{itemize}
	Based on these observations, there is a general dependence of nonlocality on Atwood number.
	Particularly, this Atwood dependence on non-locality should be considered when modeling RT mixing.
	This will be examined more closely in later sections.
}

\begin{table}
    \centering
    \begin{tabular}{c|c|c|c|c}
         Ratio & $A=0.05$ &  $A=0.3$  &  $A=0.5$ &  $A=0.8$ \\
         \hline
         $D^{10}\bigpar{h_bD^{00}}^{-1}$ & $0.11$ & $0.15$ &  $0.15$ & $0.22$ \\
         $D^{01}\bigpar{tD^{00}}^{-1}$ & $0.15$ &  $0.27$ & $0.36$ & $0.52$ \\
         $D^{20}\bigpar{h_b^2D^{00}}^{-1}$ & $0.02$ &  $0.03$ & $0.03$ & $0.06$ \\
         $D^{11}\bigpar{h_btD^{00}}^{-1}$ & $0.03$ &  $0.09$ & $0.12$ & $0.27$ \\
         $D^{02}\bigpar{t^2D^{00}}^{-1}$ & $0.02$ &  $0.03$ & $0.05$ & $0.10$ 
    \end{tabular}
    \caption{Ratios of maximum magnitudes of higher-order moments (normalized as in Figure \ref{fig:moments}) to maximum magnitudes of leading-order moments for each $A$ case.}
    \label{tab:rel_moments}
\end{table}

\begin{figure}
    \centering
    \begin{subfigure}[]{0.7\textwidth}
        \includegraphics[width=\textwidth]{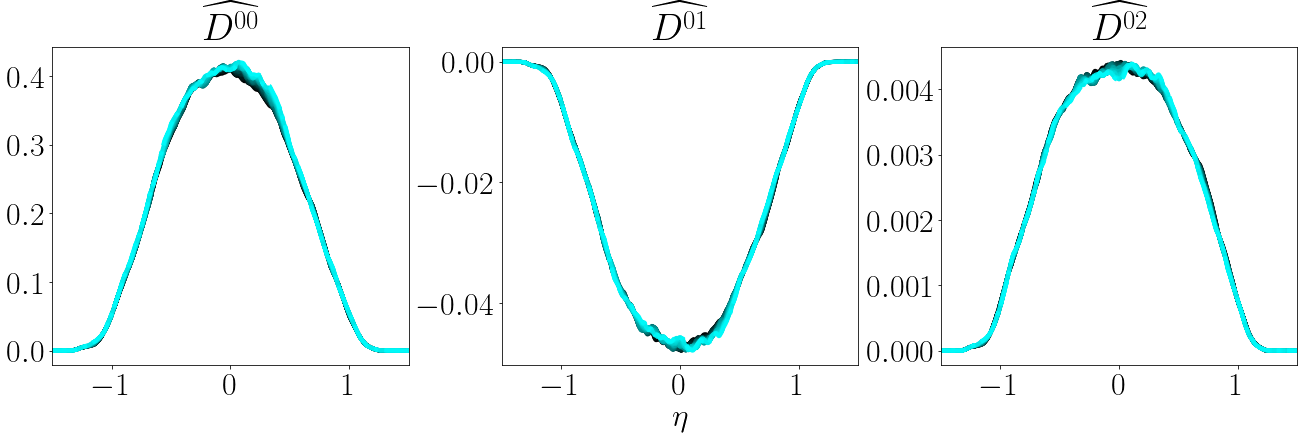}
        \subcaption[]{$A=0.05$}
        \label{subfig:selfsimtime_A005}
    \end{subfigure}
    \begin{subfigure}[]{0.7\textwidth}
        \includegraphics[width=\textwidth]{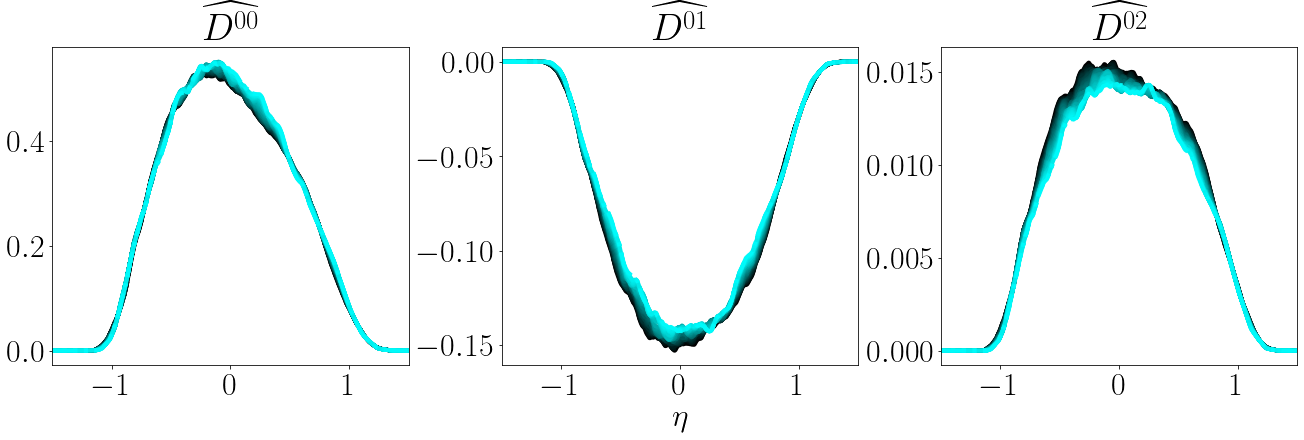} 
        \subcaption[]{$A=0.3$}
        \label{subfig:selfsimtime_A03}
    \end{subfigure}
    \begin{subfigure}[]{0.7\textwidth}
        \includegraphics[width=\textwidth]{images/moments_selfsimtime_A0.5.png}
        \subcaption[]{$A=0.5$}
        \label{subfig:selfsimtime_A05}
    \end{subfigure}
    \begin{subfigure}[]{0.7\textwidth}
        \includegraphics[width=\textwidth]{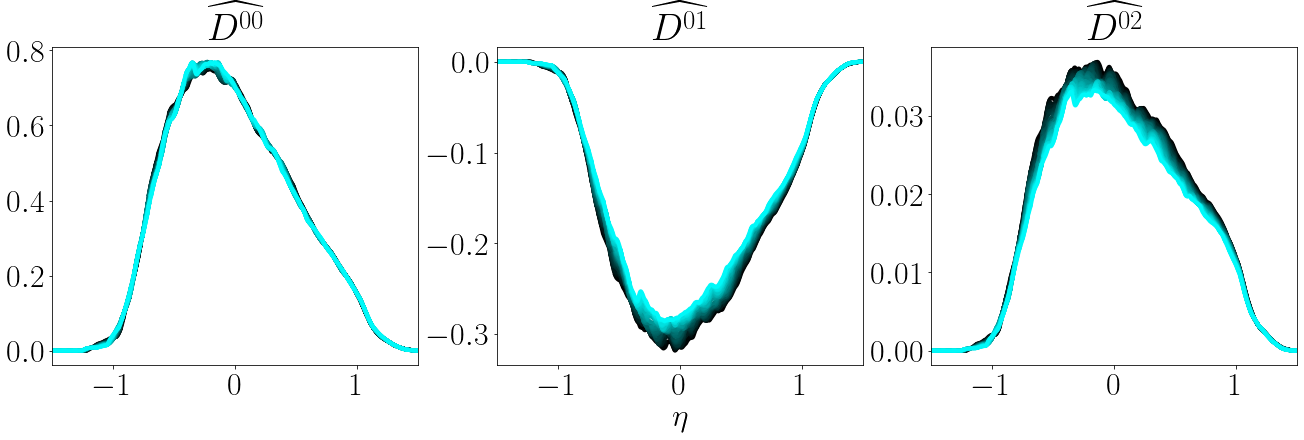}
        \subcaption[]{$A=0.8$}
        \label{subfig:selfsimtime_A08}
    \end{subfigure}
    \caption{Self-similar collapse of leading-order and higher-order temporal eddy diffusivity moments of RT instability at different Atwood numbers.}
    \label{fig:selfsim_moments_time}
\end{figure}

Here, the self-similarity of the eddy diffusivity moments is also examined.
Figure \ref{fig:selfsim_moments_time} shows the temporal eddy diffusivity moments at each Atwood number normalized according to self-similarity as in the Equations \ref{eq:selfsim1}-\ref{eq:selfsim2}; the self-similar collapse of all moments are in the Appendix in Figures \ref{fig:selfsim_A005}-\ref{fig:selfsim_A08}.
Qualitatively, the higher-order moments do not collapse as well as lower-order moments.
Additionally, the self-similar collapse worsens with increasing Atwood number.

\begin{figure}
    \centering
    \begin{subfigure}[]{0.3\textwidth}
        \includegraphics[width=\textwidth]{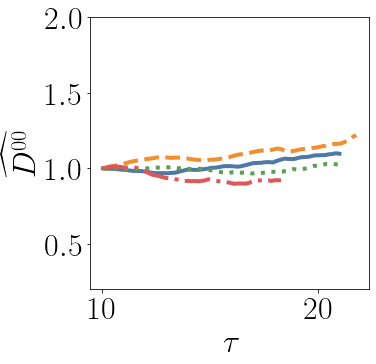}
        \subcaption[]{}
        \label{subfig:D00vt}
    \end{subfigure}
    \begin{subfigure}[]{0.3\textwidth}
        \includegraphics[width=\textwidth]{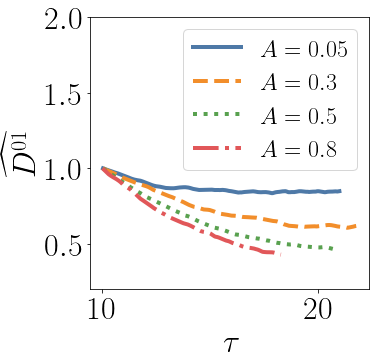}
        \subcaption[]{}
        \label{subfig:D01vt}
    \end{subfigure}
    \caption{Maximum magnitudes of normalized eddy diffusivity moments over time. Plotting starts after the time the critical $Re_T$ and $Re_L$ for turbulence are reached. Data are normalized by the values at the beginning of plotting for each Atwood case.} 
    \label{fig:selfsim_moments_v_time}
\end{figure}

The self-similarity of the moments can further be evaluated by examining the maximum magnitudes of the normalized eddy diffusivity moments.
Figure \ref{fig:selfsim_moments_v_time} shows $\widehat{D^{00}}$ and $\widehat{D^{01}}$ over the time period after which $Re_T$ exceeds $100$ and $Re_L$ exceeds $10^4$ in each Atwood case.
If these criteria are sufficient for self-similarity of the eddy diffusivity moments, the plots of the normalized moments are expected to be constant with time.
This appears to be the case for the lowest Atwood number simulation ($A=0.05$).
The higher Atwood number simulations ($A=0.5$ and $A=0.8$), however, give $\widehat{D^{00}}$ and $\widehat{D^{01}}$ that still vary in time.

Altogether, these observations suggest that higher-order moments take longer  to converge to a self-similar state than lower-order moments.
Additionally, higher-order moments take longer to reach self-similarity than lower-order quantities like the mixing width and $F$.
Thus, even if the flow in the MFM donor simulation fulfills criteria for self-similarity, such as reaching the critical Reynolds numbers or achieving a convergent $\alpha$, the eddy diffusivity moments, especially the higher-order moments, may not necessarily be self-similar.
When performing analysis on eddy diffusivity moments, one must be careful then to not only check the traditional self-similarity metrics of RT but also the self-similarity of the moments themselves.

\edit{
	Based on this, for the higher-Atwood cases, the higher-order moments are not far into the self-similar regime.
	The following analysis is performed in self-similar space in the following sections, and it is recognized that there will be some error due to the weak self-similarity of the higher-order moments.
}

%
%

\subsection{Nonlocal length and time scales}
\label{subsec:nonlocalscales}

Measurement of the eddy diffusivity moments using MFM allows for the quantification of nonlocal length and time scales.
These are defined nondimensionally as
\begin{align} 
	\eta_{NL} = \frac{1}{h_b}\sqrt{\frac{D^{20}}{D^{00}}}, \quad
	\tau_{NL} = -\frac{1}{t}\frac{D^{01}}{D^{00}}.
\end{align}

\begin{figure}[]
    \centering
    \begin{subfigure}[]{0.24\textwidth}
        \includegraphics[width=\textwidth]{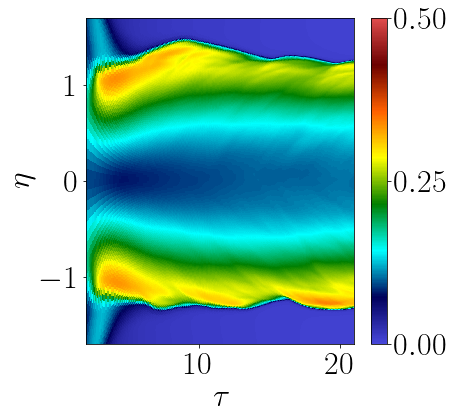}
        \subcaption[]{$A=0.05$}
        \label{subfig:etaNL_A005}
    \end{subfigure}
    \begin{subfigure}[]{0.24\textwidth}
        \includegraphics[width=\textwidth]{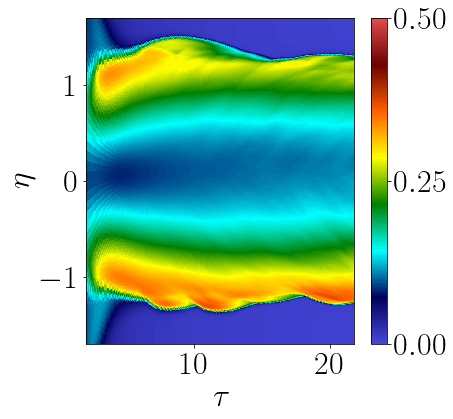}
        \subcaption[]{$A=0.3$}
        \label{subfig:etaNL_A03}
    \end{subfigure}
    \begin{subfigure}[]{0.24\textwidth}
        \includegraphics[width=\textwidth]{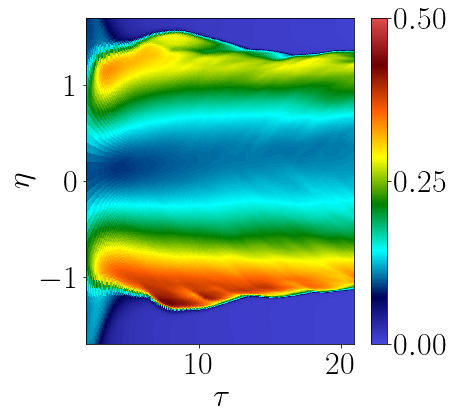}
        \subcaption[]{$A=0.5$}
        \label{subfig:etaNL_A05}
    \end{subfigure}
    \begin{subfigure}[]{0.24\textwidth}
        \includegraphics[width=\textwidth]{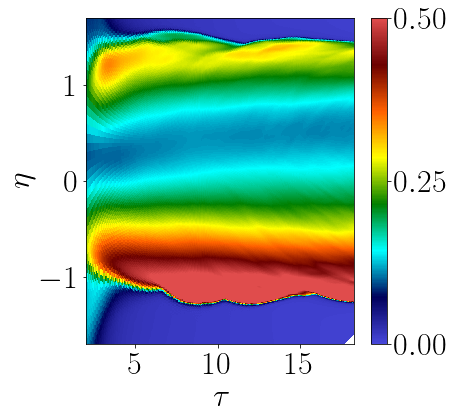}
        \subcaption[]{$A=0.8$}
        \label{subfig:etaNL_A08}
    \end{subfigure}
    \caption{Contours of nonlocal length scales for each $A$ case.}
    \label{fig:eta_NL}
\end{figure}
\begin{figure}[]
    \centering
    \begin{subfigure}[]{0.24\textwidth}
        \includegraphics[width=\textwidth]{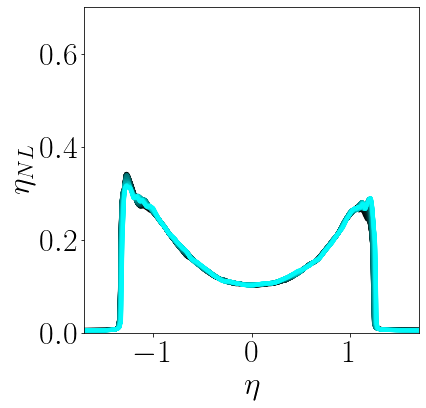}
        \subcaption[]{$A=0.05$}
        \label{subfig:etaNL_collapse_A005}
    \end{subfigure}
    \begin{subfigure}[]{0.24\textwidth}
        \includegraphics[width=\textwidth]{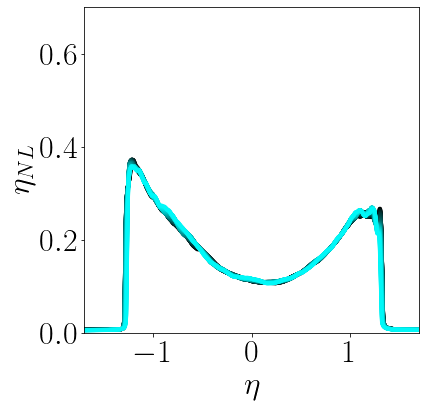}
        \subcaption[]{$A=0.3$}
        \label{subfig:etaNL_collapse_A03}
    \end{subfigure}
    \begin{subfigure}[]{0.24\textwidth}
        \includegraphics[width=\textwidth]{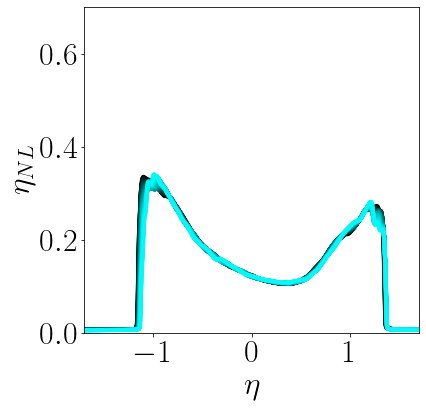}
        \subcaption[]{$A=0.5$}
        \label{subfig:etaNL_collapse_A05}
    \end{subfigure}
    \begin{subfigure}[]{0.24\textwidth}
        \includegraphics[width=\textwidth]{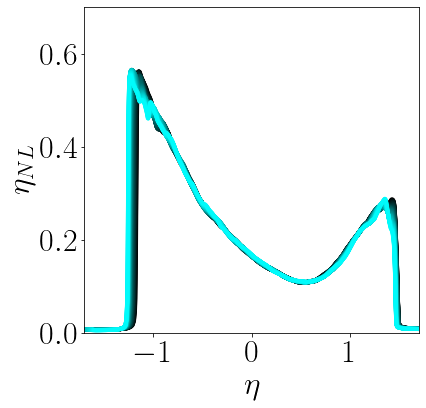}
        \subcaption[]{$A=0.8$}
        \label{subfig:etaNL_collapse_A08}
    \end{subfigure}
    \caption{Profiles of nonlocal length scales for each $A$ case. Darker lines are earlier times; lighter lines are later times.}
    \label{fig:eta_NL_collapse}
\end{figure}

Figure \ref{fig:eta_NL} shows the nonlocal length scale contours for each of the $A$ cases.
Qualitatively, they look similar across $A$, with minimum values at the centerline and maximum values at the edges of the mixing layer.
Unsurprisingly, there is increased asymmetry at higher $A$, with mixing layer edge values below the centerline greater than those above the centerline.
In the self-similar regime, profiles of the nonlocal length scales in Figure \ref{fig:eta_NL_collapse} show maximum values of approximately $\eta_{NL}=0.35$, $\eta_{NL}=0.37$, $\eta_{NL}=0.35$, and $\eta_{NL}=0.57$ for $A=0.05$, $A=0.3$, $A=0.5$, and $A=0.8$, respectively.
The minimum $\eta_{NL}$ for all $A$ is around 0.1.
Based on these observations, the following statements can be made about spatial nonlocality for late-time RT: 
\begin{itemize}
	\item $F$ at a location near the mixing layer edge depends on gradients further away from that location than does the flux at the centerline.
	\edit{
		Since mixing spreads outward, information generally propagates from the center towards the edges of the mixing layer, not the other way around.
		In this way, mixing at the edges is linked to the mean scalar gradients at the center through the flow's time history, while mixing at the center is not strongly dependent on gradients at the edges.
	}
	\item For $\eta$ at the mixing layer edges, $F$ depends on gradients approximately $0.3-0.6$ mixing half-widths away, and this value increases with $A$.
	\item For $\eta$ at the centerline, $F$ depends on gradients approximately $0.1$ mixing half-widths away, and this appears to be $A$-independent.
\end{itemize}

\begin{figure}[]
    \centering
    \begin{subfigure}[]{0.24\textwidth}
        \includegraphics[width=\textwidth]{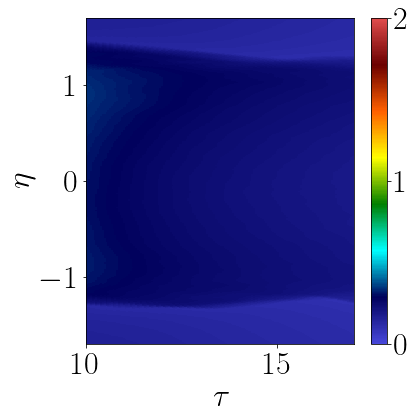}
        \subcaption[]{$A=0.05$}
        \label{subfig:tauNL_A005}
    \end{subfigure}
    \begin{subfigure}[]{0.24\textwidth}
        \includegraphics[width=\textwidth]{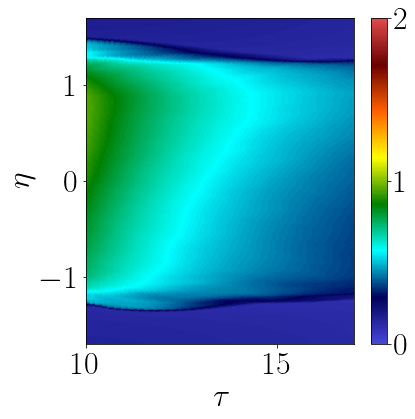}
        \subcaption[]{$A=0.3$}
        \label{subfig:tauNL_A03}
    \end{subfigure}
    \begin{subfigure}[]{0.24\textwidth}
        \includegraphics[width=\textwidth]{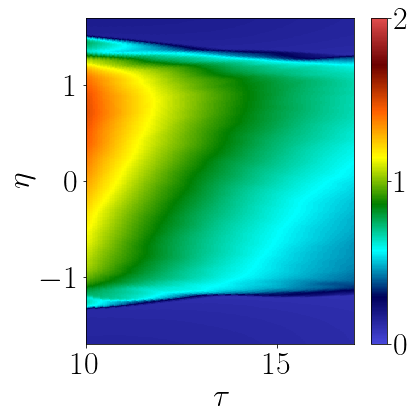}
        \subcaption[]{$A=0.5$}
        \label{subfig:tauNL_A05}
    \end{subfigure}
    \begin{subfigure}[]{0.24\textwidth}
        \includegraphics[width=\textwidth]{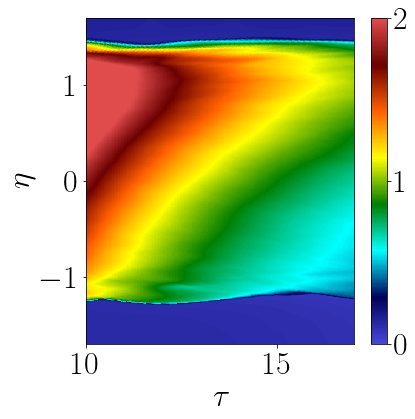}
        \subcaption[]{$A=0.8$}
        \label{subfig:tauNL_A08}
    \end{subfigure}
    \caption{Contours of nonlocal time scales for each $A$ case.}
    \label{fig:tau_NL}
\end{figure}

\begin{figure}[]
    \centering
    \begin{subfigure}[]{0.24\textwidth}
        \includegraphics[width=\textwidth]{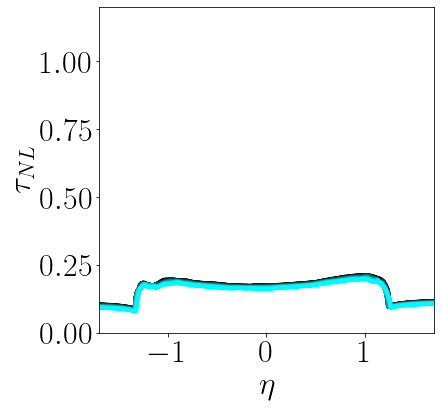}
        \subcaption[]{$A=0.05$}
        \label{subfig:tauNL_collapse_A005}
    \end{subfigure}
    \begin{subfigure}[]{0.24\textwidth}
        \includegraphics[width=\textwidth]{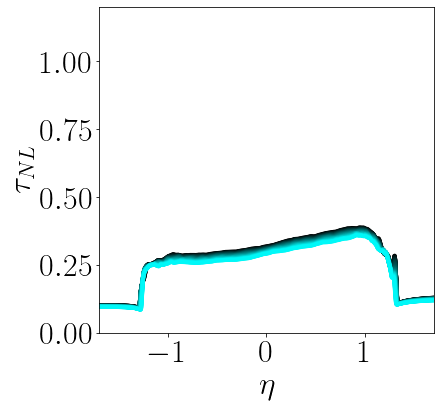}
        \subcaption[]{$A=0.3$}
        \label{subfig:tauNL_collapse_A03}
    \end{subfigure}
    \begin{subfigure}[]{0.24\textwidth}
        \includegraphics[width=\textwidth]{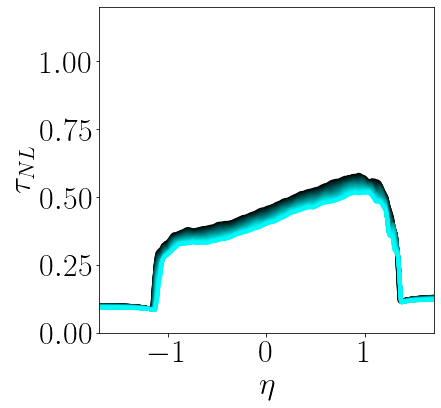}
        \subcaption[]{$A=0.5$}
        \label{subfig:tauNL_collapse_A05}
    \end{subfigure}
    \begin{subfigure}[]{0.24\textwidth}
        \includegraphics[width=\textwidth]{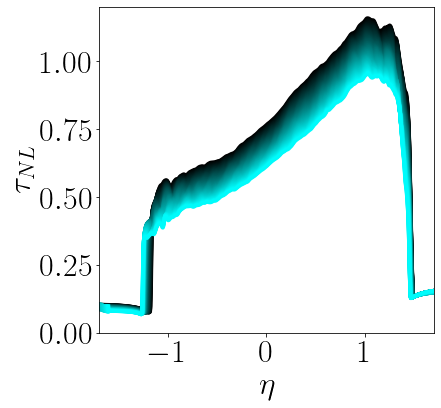}
        \subcaption[]{$A=0.8$}
        \label{subfig:tauNL_collapse_A08}
    \end{subfigure}
    \caption{Profiles of nonlocal time scales scaled by $\tau$ for each $A$ case. Darker lines are earlier times; lighter lines are later times.}
    \label{fig:tau_NL_collapse}
\end{figure}

The nonlocal time scales are also examined in Figure \ref{fig:tau_NL}.
In contrast to the nonlocal length scale, the nonlocal time scale differs greatly over the $A$ studied here.
Particularly, the max values of $\tau_{NL}$ increase with $A$, indicating that $F$ depends more on earlier times for higher $A$.
Additionally, the contours for $\tau_{NL}$ become more asymmetric with increasing $A$---max $\tau_{NL}$ shifts towards the edge of the mixing layer above the centerline as $A$ increases.
It is expected that $\tau_{NL}$ profiles collapse in the self-similar unit, which is inspected for each $A$ case in Figure \ref{fig:tau_NL_collapse}.
It must be noted that the quality of the collapse worsens as $A$ increases, indicating that our highest $A$ cases may not be far into the self-similar regime.
Nevertheless, the profiles there show maximum values of approximately $\tau_{NL}=0.21\tau$, $\tau_{NL}=0.39\tau$, $\tau_{NL}=0.58\tau$, and $\tau_{NL}=1.16\tau$ for $A=0.05$, $A=0.3$, $A=0.5$, and $A=0.8$, respectively.
Based on these observations, the following statements can be made about temporal nonlocality for late-time RT: 
\begin{itemize}
	\item As $A$ increases, across the mixing layer, $F$ depends more on the flux at earlier times.
	\item At low $A$, the dependence of $F$ on earlier times is relatively uniform across the mixing layer.
	\item As $A$ increases, $F$ near the upper edge of the mixing layer depends on earlier times than does the flux at the lower edge. 
	\edit{
		This suggests that, compared to the spikes, it may take the bubbles a longer time to transition to a self-similar state where the flow forgets its initial conditions, and this effect appears stronger at higher $A$.
	}
\end{itemize}

The MFM measurements also reveal large $\tau_{NL}$ at early times across $A$.
These high $\tau_{NL}$ zones appear to be higher in magnitude and last longer as $A$ increases.
This suggests that as $A$ increases, the RT instability retains memory of the initial conditions for a longer period of time.

\begin{figure}
	\centering
	\includegraphics[width=0.25\textwidth]{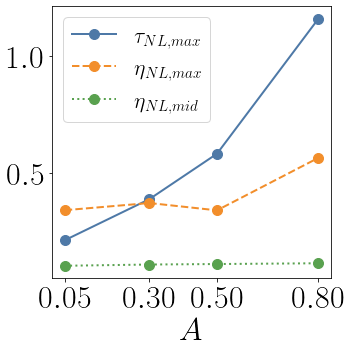}
	\caption{Maximum nonlocal time scale, maximum nonlocal length scale, and nonlocal length scale at centerline over $A$.}
	\label{fig:nonlocal_scales}
\end{figure}

Figure \ref{fig:nonlocal_scales} shows the maximum $\tau_{NL}$ and $\eta_{NL}$ and the centerline $\eta_{NL}$ over the studied $A$.
The centerline $\eta_{NL}$ do not change much with $A$.
The maximum $\eta_{NL}$ appears to have some sensitivity to $A$ and potentially increases at high $A$, but this is difficult to discern with the limited data.
On the other hand, the maximum $\tau_{NL}$ increases rapidly with $A$.
This indicates that while there is some dependence of spatial nonlocality on $A$, the dependence of temporal nonlocality on $A$ is larger and more important for modeling RT mixing.
Such a dependence should be captured by RANS models for accurate prediction of mixing due to RT instability.

\subsection{Kramers-Moyal terms}
\label{subsec:km}

\begin{figure}
    \centering
    \begin{subfigure}[]{0.24\textwidth}
        \includegraphics[height=9em]{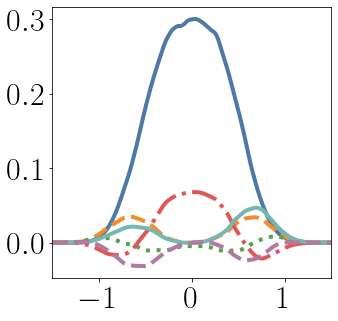}
        \subcaption[]{$A=0.05$}
        \label{subfig:KM_A005}
    \end{subfigure}
    \begin{subfigure}[]{0.24\textwidth}
        \includegraphics[height=9em]{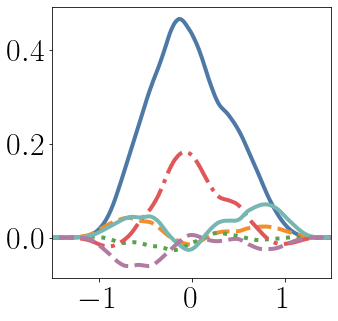}
        \subcaption[]{$A=0.3$}
        \label{subfig:KM_A03}
    \end{subfigure}
    \begin{subfigure}[]{0.24\textwidth}
        \includegraphics[height=9em]{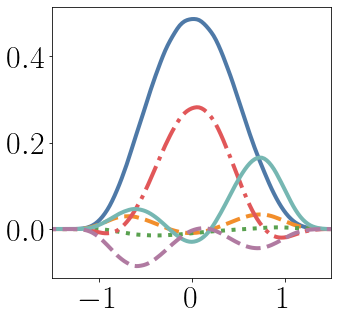}
        \subcaption[]{$A=0.5$}
        \label{subfig:KM_A05}
    \end{subfigure}
    \begin{subfigure}[]{0.24\textwidth}
        \includegraphics[height=9em]{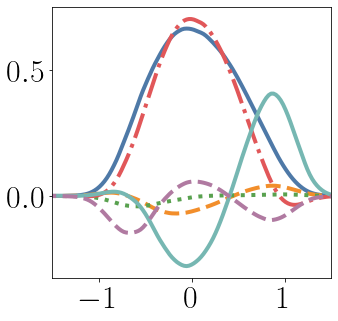}
        \subcaption[]{$A=0.8$}
        \label{subfig:KM_A08}
    \end{subfigure}
    \caption{Terms of the Kramers-Moyal expansion for $F$ at different Atwood numbers. 
    Each line corresponds to terms with contributions from: $D^{00}$ (solid blue), $D^{10}$ (dashed orange), $D^{01}$ (dash-dotted red), $D^{20}$ (dotted green), $D^{11}$ (solid teal), and $D^{02}$ (dashed lilac).}
    \label{fig:kmterms}
\end{figure}

To further assess the importance of nonlocality, the terms in the Kramers-Moyal expansion (Equation \ref{eq:KME}) for $F$ can be examined, as shown in Figure \ref{fig:kmterms}.
These terms are calculated a priori: the donor simulation $\YHm$ is used for the mean mass fraction gradients, and the measured eddy diffusivity moments are substituted directly.
Already at the lowest Atwood case of $A=0.05$, the higher-order terms appear non-negligible compared to the leading-order term; at least some of the higher-order terms will need to be retained for complete characterization of the eddy diffusivity.
This was also shown in the 2D case at the same Atwood number studied in \citet{lavacot2024}.
Furthermore, as Atwood number increases, the higher-order terms become closer in magnitude to the leading-order term, indicating that nonlocality becomes more important with increasing Atwood.
This also suggests that at higher Atwood numbers, more higher-order moments may be required for modeling than at lower Atwood numbers.
It is notable that the temporal moments are particularly large at high Atwood numbers, indicating that temporal nonlocality may be especially important in those regimes.

\section{Importance of moments in modeling} 
\label{sec:modeling}
The previous section focused on assessment of nonlocality through the measurement of eddy diffusivity moments using MFM.
While this processes revealed the importance of nonlocality, it has not yet been shown which of the eddy diffusivity moments are important for modeling.
By testing different combinations of moments in a model form, the moments most important for modeling can be determined, and it can be discerned whether this depends on Atwood number.
Here, an \textit{inverse operator} is proposed to incorporate information about nonlocality from the eddy diffusivity moments.

\subsection{Matched Moment Inverse}
\label{subsec:MMI}

Truncation of Equation \ref{eq:KME} represents an approximate model for the turbulent species flux. 
In fact, truncation to the first term is the gradient-diffusion approximation.
However, a property of the Kramers-Moyal expansion is that it does not converge with finite terms, so adding higher-order terms to the leading-order term can lead to divergence \cite{pawula1967}.

Instead, the Matched Moment Inverse (MMI) \cite{liu2023}, a systematic method for constructing a model using eddy diffusivity moments, is employed.
With MMI, the goal is to match the shape of the eddy diffusivity kernel using its moments.
This is achieved by determining coefficients $a^{mn}(y,t)$ for the \textit{inverse operator}:
\begin{align}
	\bigbra{1 + a^{10}\diffp{}{y} + a^{01}\diffp{}{t} + a^{20}\diffp{}{{y^2}}+...}F = a^{00}\rhom\diffp{\YHm}{y}.
	\label{eq:MMI_form_phys}
\end{align}
The inverse operator on the left hand side can be expanded based on which moments are used; $a^{mn}$ corresponds to using $D^{mn}$.
The model coefficients are determined numerically using MFM simulation data.
For example, if $D^{00}$, $D^{01}$, $D^{10}$, and $D^{20}$ are used, the following system is solved for $a^{00}$, $a^{01}$, $a^{10}$, and $a^{20}$:
\begin{align}
	\left[1+a^{10}\frac{\partial}{\partial y}+a^{01}\frac{\partial}{\partial t}+a^{20}\frac{\partial^2}{\partial y^2}\right]F^{00}&=a^{00}\rhom,\label{eq:MMI_num_1}\\
	\left[1+a^{10}\frac{\partial}{\partial y}+a^{01}\frac{\partial}{\partial t}+a^{20}\frac{\partial^2}{\partial y^2}\right]F^{10}&=a^{00} \rhom \left(y-\frac{1}{2}\right),\\
	\left[1+a^{10}\frac{\partial}{\partial y}+a^{01}\frac{\partial}{\partial t}+a^{20}\frac{\partial^2}{\partial y^2}\right]F^{01}&=a^{00}\rhom t,\\
	\left[1+a^{10}\frac{\partial}{\partial y}+a^{01}\frac{\partial}{\partial t}+a^{20}\frac{\partial^2}{\partial y^2}\right]F^{20}&=a^{00}\rhom \frac{1}{2}\left(y-\frac{1}{2}\right)^2,\label{eq:MMI_num_2}
\end{align}
where $F^{mn}$ is the post-processed turbulent species flux resulting from a macroscopic forcing achieving $\frac{\partial\Ycm}{\partial y}=(y-\frac{1}{2})^mt^n$ to determine $D^{mn}$.

The above is demonstrated for spatio-temporal variables for simplicity, but the analysis presented here is done in self-similar space.
The self-similar inverse operator is
\begin{align}
	\bigbra{1 
		+ \widehat{a^{10}}\diffp{}{\eta} 
		+ \widehat{a^{01}}\bigpar{1-2\eta\diffp{}{\eta} }
		+ \widehat{a^{20}}\diffp{}{{\eta^2}}
		+...}\widehat{F} &= \widehat{a^{00}}\rhom\diffp{\YHm}{\eta},
	\label{eq:MMI_form_ss}    
\end{align}
where the self-similar coefficients are

\begin{align}
	\widehat{a^{00}}&=\frac{1}{{\alpha^*}^2 A^2g^2(t-t^*)^3}a^{00},\\
	\widehat{a^{01}}&=\frac{1}{t-t^*}a^{01},\\
	\widehat{a^{10}}&=\frac{1}{\alpha^* Ag(t-t^*)^2}a^{10},\\
	\widehat{a^{20}}&=\frac{1}{{\alpha^*}^2 A^2g^2(t-t^*)^4}a^{20}.
\end{align}


Since $D^{mn}$ are taken directly from MFM measurements, they contain some statistical error, which would be amplified by the MMI fitting process and obfuscate analysis.
To avoid this, a moving average filter is applied to the moments, and those filtered moments are used for the inverse operator coefficient fitting process, done by solving Equations \ref{eq:MMI_num_1}-\ref{eq:MMI_num_2}.
Equation \ref{eq:MMI_form_phys} is then solved using $\rhom$ and $\YHm$ obtained from the simulations.
In the following sections, coefficients in Equation \ref{eq:MMI_form_phys} are fit in self-similar space using different combinations of $D^{mn}$ to be tested in the inverse operator.

\subsubsection{Inverse operator using only spatial moments of eddy diffusivity}

\begin{figure}
    \centering
    \begin{subfigure}[]{0.24\textwidth}
        \includegraphics[height=9em]{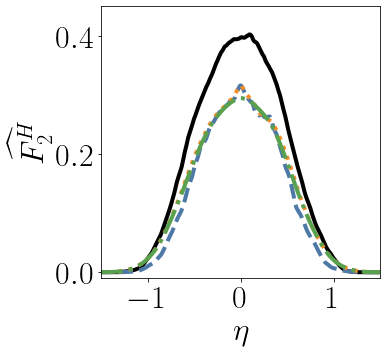}
        \subcaption[]{$A=0.05$}
        \label{subfig:MMI_spatial_A005}
    \end{subfigure}
    \begin{subfigure}[]{0.24\textwidth}
        \includegraphics[height=9em]{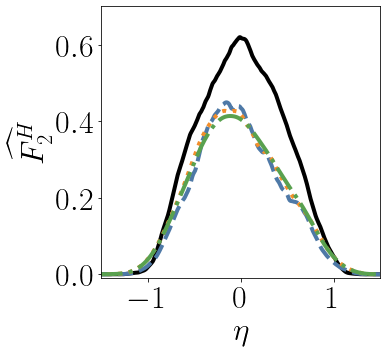}
        \subcaption[]{$A=0.3$}
        \label{subfig:MMI_spatial_A03}
    \end{subfigure}
    \begin{subfigure}[]{0.24\textwidth}
        \includegraphics[height=9em]{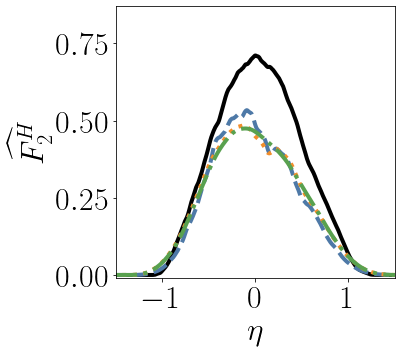}
        \subcaption[]{$A=0.5$}
        \label{subfig:MMI_spatial_A05}
    \end{subfigure}
    \begin{subfigure}[]{0.24\textwidth}
        \includegraphics[height=9em]{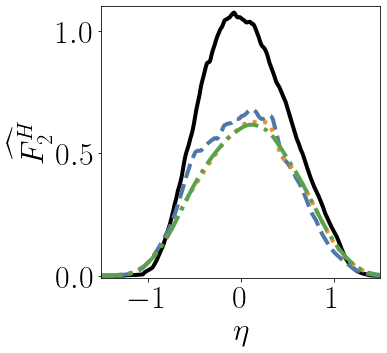}
        \subcaption[]{$A=0.8$}
        \label{subfig:MMI_spatial_A08}
    \end{subfigure}
    \caption{$F$ predicted using a closure based on spatial moments. The solid black line is $F$ computed from the high-fidelity simulation. The other lines are predictions from model forms using different combinations of moments, as follows: $D^{00}$ (dashed blue); $D^{00}$ and $D^{10}$ (dotted orange); and $D^{00}$, $D^{10}$, and $D^{20}$ (dash-dotted green).}
    \label{fig:F_spatial}
\end{figure}

The inverse operator using only spatial moments is tested first.
In physical space, the model forms are
\begin{align}
    F = a^{00}\rhom\diffp{\YHm}{y},\\
    \bigbra{1 + a^{10}\diffp{}{y} }F = a^{00}\rhom\diffp{\YHm}{y},\\
    \bigbra{1 + a^{10}\diffp{}{y}  + a^{20}\diffp{}{{y^2}} }F = a^{00}\rhom\diffp{\YHm}{y}.
    \label{eq:MMI_spatial}
\end{align}
Note the first model form uses only the leading-order moment and is equivalent to the truncation of Equation \ref{eq:KME} to the leading-order term.
Figure \ref{fig:F_spatial} shows $F$ predicted using the above model for each Atwood case.
In these figures, the solid black lines are the self-similar $\widehat{F}$ taken from the last timesteps of the donor simulations shown in Figure \ref{fig:tsf_selfsim}.
Across all $A$,  there is little improvement when higher-order spatial moments are added.
The biggest change with the addition of higher-order moments is at the edges of the mixing layer, as the width of the predicted $F$ becomes closer to $F$ from the high-fidelity simulation---this is most obvious in the $A=0.05$ case.
This indicates that, while spatial nonlocality is non-negligible, most improvements will come from the temporal moments.

\subsubsection{Inverse operator using spatial and temporal moments of eddy diffusivity}

\begin{figure}
    \centering
    \begin{subfigure}[]{0.24\textwidth}
        \includegraphics[height=9em]{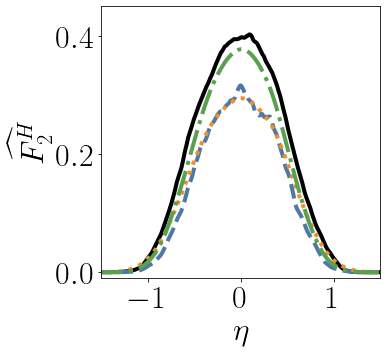}
        \subcaption[]{$A=0.05$}
        \label{subfig:MMI_temporal_A005}
    \end{subfigure}
    \begin{subfigure}[]{0.24\textwidth}
        \includegraphics[height=9em]{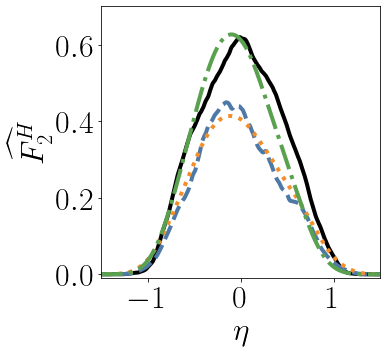}
        \subcaption[]{$A=0.3$}
        \label{subfig:MMI_temporal_A03}
    \end{subfigure}
    \begin{subfigure}[]{0.24\textwidth}
        \includegraphics[height=9em]{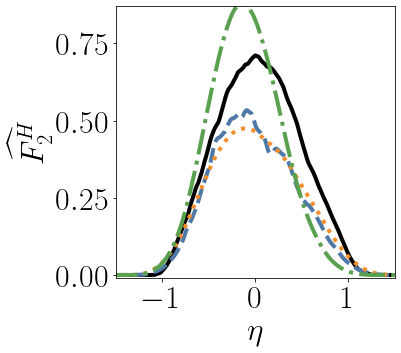}
        \subcaption[]{$A=0.5$}
        \label{subfig:MMI_temporal_A05}
    \end{subfigure}
    \begin{subfigure}[]{0.24\textwidth}
        \includegraphics[height=9em]{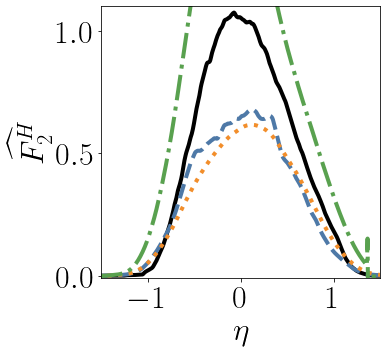}
        \subcaption[]{$A=0.8$}
        \label{subfig:MMI_temporal_A08}
    \end{subfigure}
    \caption{$F$ predicted using a closure based on spatial and temporal moments. The solid black line is $F$ computed from the high-fidelity simulation. The other lines are predictions from operators using different combinations of moments, as follows: $D^{00}$ (dashed blue); $D^{00}$, $D^{10}$, and $D^{20}$ (dotted orange); and $D^{00}$, $D^{01}$, $D^{10}$, and $D^{20}$ (dash-dotted green).}
    \label{fig:F_temporal}
\end{figure}

An inverse operator using $D^{00}$, $D^{10}$, $D^{20}$, and $D^{01}$ is now assessed.
\begin{align}
    \bigbra{1 + a^{10}\diffp{}{y}  + a^{20}\diffp{}{{y^2}} + a^{01}\diffp{}{t} }F = a^{00}\rhom\diffp{\YHm}{y}.
    \label{eq:MMI_D00_D10_D01_D20}
\end{align}

Figure \ref{fig:F_temporal} shows predictions for $F$ using this spatio-temporal operator.
There is a larger change in $F$ prediction when adding the first temporal moment than higher-order spatial moments.
Particularly, for $A=0.05$, $F$ prediction is significantly improved by addition of $D^{01}$ to the inverse operator and matches $F$ computed from the high-fideltiy simulation well in both magnitude and shape.
Additionally, there is a marked improvement in the prediction for the $A=0.3$ case, though it does not match the shape as well.
While improvements are observed in $F$ predictions in these cases, there is diverging behavior in the higher Atwood cases of $0.5$ and $0.8$.
This is because, for these Atwood cases, the MMI fitting process gives coefficients of intuitively incorrect signs, resulting in operators that are not robust.
The requirements on the MMI coefficients for model robustness can be found by rearranging Equation \ref{eq:MMI_D00_D10_D01_D20}:
\begin{align}
    \diffp{}{t} F = -\frac{1}{a^{01}}F - \frac{a^{10}}{a^{01}}\diffp{}{y}F  - \frac{a^{20}}{a^{01}}\diffp{}{{y^2}}F + \frac{a^{00}}{a^{01}}\rhom\diffp{\YHm}{y}.
    \label{eq:MMI_D00_D10_D01_D20_transp}
\end{align}
The first term on the right hand side must be a destruction term for stability.
Thus, $a^{01}$ must be always positive for robustness.
As demonstrated in Appendix \ref{sec:model_robustness}, this is achieved when $-\frac{\widehat{D^{01}}}{\widehat{D^{00}}}<0.25$.

\begin{figure}[t]
	\centering
	\includegraphics[width=0.3\textwidth]{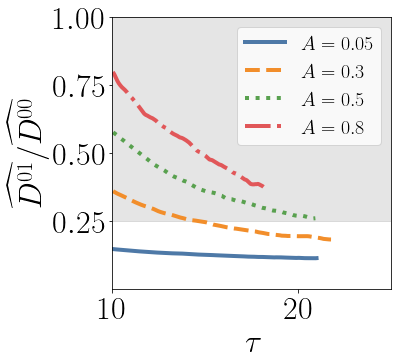}
	\caption{The maximum $-\frac{\widehat{D^{01}}}{\widehat{D^{00}}}$ over $\eta$ within the mixing zone as a funciton of time for each Atwood case. Plotting starts after the time the critical $Re_T$ and $Re_L$ for turbulence are reached. }
	\label{fig:D01_D00_ratio}
\end{figure}

Figure \ref{fig:D01_D00_ratio} shows the maximum $-\frac{\widehat{D^{01}}}{\widehat{D^{00}}}$ plotted over time for each of the Atwood numbers studied.
In the self-similar regime, $-\frac{\widehat{D^{01}}}{\widehat{D^{00}}}$ is expected to be constant with $\tau$.
Indeed, the $A=0.05$ and $A=0.3$ cases both result in relatively constant $-\frac{\widehat{D^{01}}}{\widehat{D^{00}}}$ at late $\tau$ that goes under $0.25$ before the end of the simulation, which explains why those cases give robust models after the MMI process.
Errors in Figure \ref{fig:F_temporal} are maybe due to fitting the inverse operator coefficients to eddy diffusivity moments that are not yet self-similar.
Thus, the $-\frac{\widehat{D^{01}}}{\widehat{D^{00}}}$ constraint discovered here may be another indicator of self-similarity of the RT instability eddy diffusivity moments.
\edit{
	That is, though our simulations are fully-turbulent, the fact that the higher-Atwood simulations do not reach this $-\frac{\widehat{D^{01}}}{\widehat{D^{00}}}$ threshold indicates that the higher-order moments are not far into the self-similar regime, which is why the data leads to non-robust model coefficients for these cases.
}

It must be noted that while the predictions of $F$ in Figures \ref{fig:F_spatial} and \ref{fig:F_temporal} are fairly smooth, the inverse operator coefficients obtained through MMI (shown as the solid blue lines in Figure \ref{fig:model_coeffs}) contain large fluctuations over $\eta$, even exhibiting some singlularities.
This behavior of the MMI-obtained coefficients has been observed by \citet{liu2023}.
Thus, these fitted coefficients should be used only as a guide for determining model coefficients.
The following section proposes a framework for analytically representing the inverse operator coefficients for use in a model.

\subsubsection{Atwood dependence of nonlocality in a model}
\label{sec:atwood_model}

\begin{figure}
	\centering
	\begin{subfigure}[]{0.24\textwidth}
		\includegraphics[height=9em]{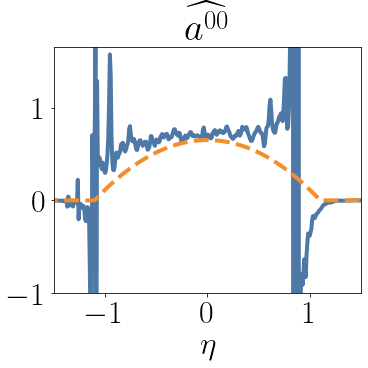}
		\subcaption[]{}
		\label{subfig:model_a00}
	\end{subfigure}
	\begin{subfigure}[]{0.24\textwidth}
		\includegraphics[height=9em]{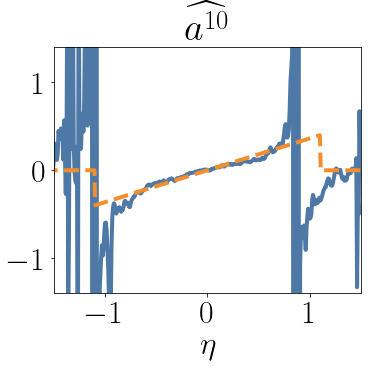}
		\subcaption[]{}
		\label{subfig:model_a10}
	\end{subfigure}
	\begin{subfigure}[]{0.24\textwidth}
		\includegraphics[height=9em]{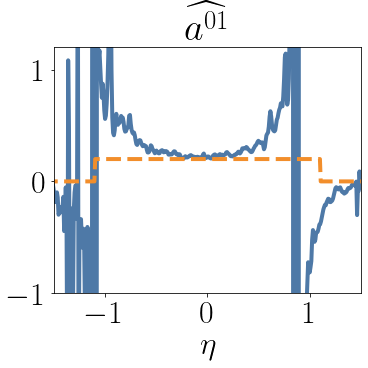}
		\subcaption[]{}
		\label{subfig:model_a01}
	\end{subfigure}1
	\begin{subfigure}[]{0.24\textwidth}
		\includegraphics[height=9em]{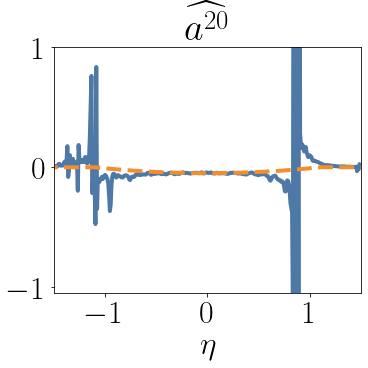}
		\subcaption[]{}
		\label{subfig:model_a20}
	\end{subfigure}
	\caption{Coefficients of proposed model (dashed orange) compared to $a^{mn}$ determined from MFM measurements (solid blue) at $A=0.05$.}
	\label{fig:model_coeffs}
\end{figure}

Here, a model is proposed using coefficients written algebraically in self-similar space based on the findings previously discussed.
\edit{
	The main intent here is to present a framework for incorporation of nonlocality and its Atwood dependence based on the MFM analysis in the previous sections, not provide a complete model to be employed as is.
	The result is a geometrically-defined model depending on $\eta$, but a complete model should employ functions of other variables in the RANS model (e.g., $k$) to allow for generality. 
	Thus, the model proposed here represents  the first steps towards incorporating Atwood-dependent non-locality informed by direct measurements into a turbulent mixing model, and future work would involve integrating these findings into a complete, usable model.
}
The \edit{algebraic} model is of the following form:
\begin{align}
	\bigbra{1 + \mathscr{a}^{10}\diff{}{\eta}  + \mathscr{a}^{20}\diff{}{{\eta^2}} + \mathscr{a}^{01}\bigpar{1-2\diff{}{\eta} }}\widehat{F} = \mathscr{a}^{00}\rhom\diff{\YHm}{\eta}.
	\label{eq:atwood_model}
\end{align}
$\mathscr{a}^{mn}$ are model coefficients associated with moments $D^{mn}$ but algebraically defined within the mixing zone:
\begin{align}
    \mathscr{a}^{00}&=\bigpar{1.4A +0.58 }\bigpar{1-\bigpar{\frac{\eta}{\xi}}^2 }^{\frac{1}{2}},\\
    \mathscr{a}^{10}&=\bigpar{0.4A +0.38} \frac{\eta}{\xi},\\
    \mathscr{a}^{01}&=0.4A+0.18,\\
    \mathscr{a}^{20}&=\bigpar{-0.05A -0.04 }\bigpar{1-\bigpar{\frac{\eta}{\xi}}^2 },
    \label{eq:atwood_model_coeffs}
\end{align}
where $\xi$ is chosen to be $1.1$, and $\mathscr{a}^{mn}$ are zero outside $\pm\xi$.
\edit{
	The functional forms of the model coefficients are chosen to match the shapes of the $a^{mn}$ obtained through the MMI procedure on the MFM measurements of the eddy diffusivity moments from the $A=0.05$ case, as shown in Figure \ref{fig:model_coeffs} .
}
The dependence on $A$ of each $\mathscr{a}^{mn}$ is determined such that the predictions of $F$ resulting from the model are close to $F$ from the high fidelity simulation.
A linear dependence on $A$ is chosen for its simplicity and appears to suit the cases considered here well.
More data at different $A$ could confirm the dependence proposed here or inform a more accurate fit.


\begin{figure}
    \centering
    \begin{subfigure}[]{0.24\textwidth}
        \includegraphics[height=9em]{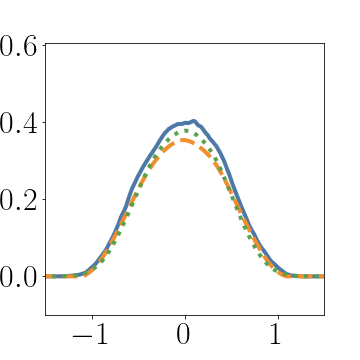}
        \subcaption[]{$A=0.05$}
        \label{subfig:detterms_A005}
    \end{subfigure}
    \begin{subfigure}[]{0.24\textwidth}
        \includegraphics[height=9em]{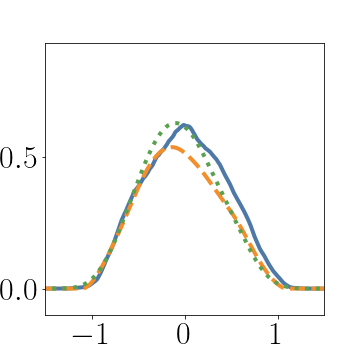}
        \subcaption[]{$A=0.3$}
        \label{subfig:dettermsA03}
    \end{subfigure}
    \begin{subfigure}[]{0.24\textwidth}
        \includegraphics[height=9em]{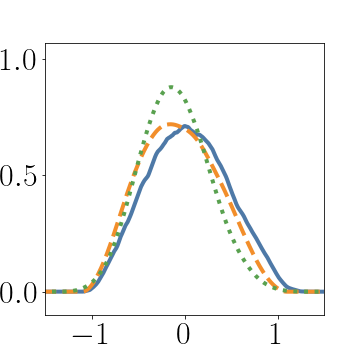}
        \subcaption[]{$A=0.5$}
        \label{subfig:detterms_A05}
    \end{subfigure}
    \begin{subfigure}[]{0.24\textwidth}
        \includegraphics[height=9em]{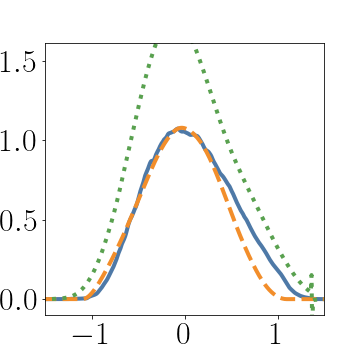}
        \subcaption[]{$A=0.8$}
        \label{subfig:detterms_A08}
    \end{subfigure}
    \caption{$F$ predictions from ensemble-averaged high fidelity simulations (solid blue), model using analytically-defined coefficients (dashed orange), and model using coefficients determined from MFM measurements (dotted green).}
    \label{fig:model_tsf}
\end{figure}

Figure \ref{fig:model_tsf} shows predictions of $F$ using the proposed model and a model using coefficients found through the MMI (henceforth referred to as the MFM-based model) procedure compared to $F$ from high fidelity simulations.
These predictions are obtained by solving the proposed model in Equation \ref{eq:atwood_model} and the MFM-based model in Equation \ref{eq:MMI_D00_D10_D01_D20} using $\rhom$ and $\YHm$ from the high fidelity simulations.
The proposed model predicts $F$ close to that from the high-fidelity simulations for all Atwood numbers.
This is an improvement from the MFM-based model, which overpredicts $F$ at higher Atwood number ($A\geq0.5$), since the simulations for those $A$ give $-\frac{\widehat{D^{01}}}{\widehat{D^{00}}}$ above the threshold of $0.25$.
This suggests that the proposed linear dependence of the model on $A$ may be sufficient for this range of $A$, but more data should be gathered to confirm whether this is the case, especially at higher $A$ not examined here.

The effective eddy diffusivity moments of the proposed model are now examined and compared with the eddy diffusivity moments measured from high fidelity simulations using MFM.
To obtain the model eddy diffusivity moments, MFM is applied directly to the model by specifying $\diff{\YHm}{\eta}$ in the self-similar MMI equations and solving them for $\widehat{F^{mn}}$.
The resulting equations are
\begin{align}
     \bigbra{1+ \widehat{a^{10}}\diff{}{\eta} +\widehat{a^{01}}\bigpar{3-2\eta\diff{}{\eta}} + \widehat{a^{20}}\diff[2]{}{\eta}}\widehat{F^{00}} &= \widehat{a^{00}}\rhom, \\
     \bigbra{1+ \widehat{a^{10}}\diff{}{\eta} +\widehat{a^{01}}\bigpar{5-2\eta\diff{}{\eta}} + \widehat{a^{20}}\diff[2]{}{\eta}}\widehat{F^{10}} &= \widehat{a^{00}}\rhom\eta, \\
     \bigbra{1+ \widehat{a^{10}}\diff{}{\eta} +\widehat{a^{01}}\bigpar{4-2\eta\diff{}{\eta}} + \widehat{a^{20}}\diff[2]{}{\eta}}\widehat{F^{01}} &= \widehat{a^{00}}\rhom, \\
     \bigbra{1+ \widehat{a^{10}}\diff{}{\eta} +\widehat{a^{01}}\bigpar{7-2\eta\diff{}{\eta}} + \widehat{a^{20}}\diff[2]{}{\eta}}\widehat{F^{20}} &= \widehat{a^{00}}\rhom\frac{\eta^2}{2}.
\end{align}
The model $\widehat{D^{mn}}$ are computed from the $\widehat{F^{mn}}$:
\begin{align}
   \widehat{D^{00}} &= \frac{\widehat{F^{00}}}{\rhom},\\
    \widehat{D^{10}} &= \frac{\widehat{F^{10}} - \eta\rhom\widehat{D^{00}}}{\rhom} ,\\
    \widehat{D^{01}} &= \frac{\widehat{F^{01}} - \rhom\widehat{D^{00}}}{\rhom} ,\\
    \widehat{D^{20}} &= \frac{\widehat{F^{20}} - \eta\rhom\widehat{D^{10}} - \frac{\eta^2}{2}\rhom\widehat{D^{00}}}{\rhom}.
\end{align}

\begin{figure}
    \centering
    \resizebox{.8\linewidth}{!}{\input{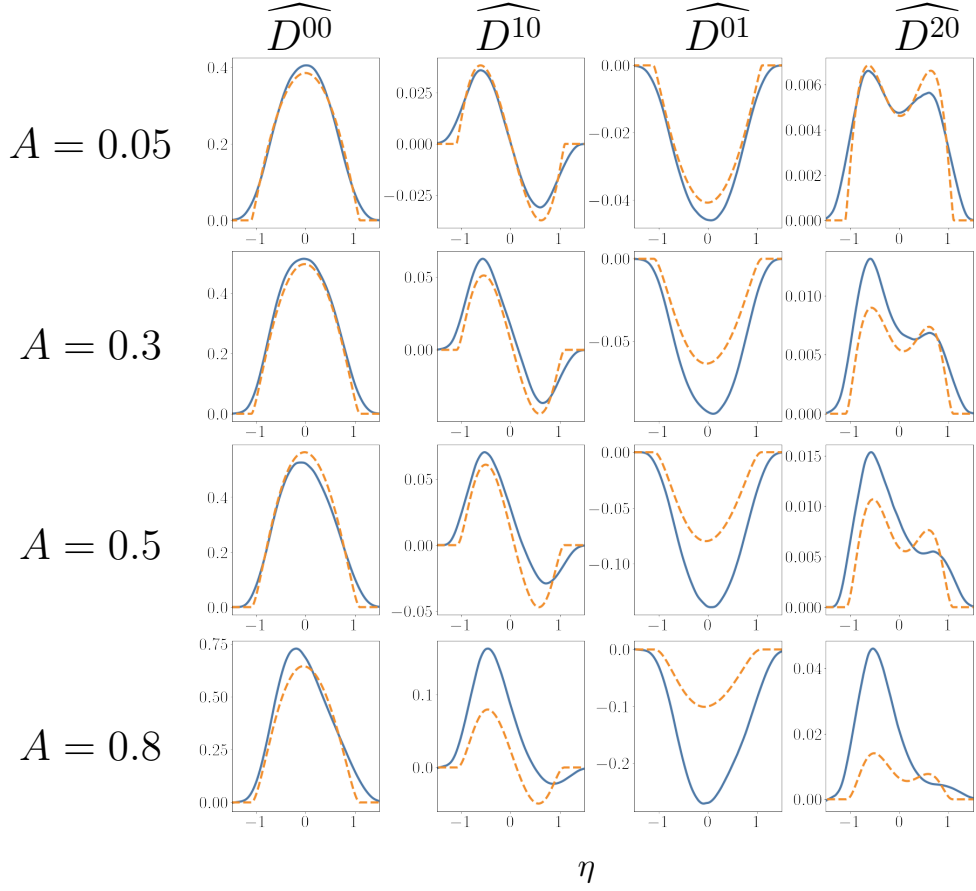}}
    \caption{Eddy diffusivity moments of proposed model (dashed orange) compared to MFM measurements (solid blue).}
    \label{fig:model_moments}
\end{figure}

Figure \ref{fig:model_moments} shows the model eddy diffusivity moments compared to the eddy diffusivity moments from high-fidelity simulations, the latter of which is now referred to as the ``true'' eddy diffusivities.
It is not surprising that the model eddy diffusivity moments at $A=0.05$ are close to the true moments, since the model coefficients $\mathscr{a}^{mn}$ are fit to that Atwood case.
Better fits could potentially be achieved by using more sophisticated functions of $\eta$ to better capture the smoothness of the profiles at the edges of the mixing layer.
As $A$ increases, the model moments are not as close to the true moments, since the 
\edit{
coefficients are not tuned to result in moments matching those measured in the high-fidelity simulations.
}
The most notable discrepancy is among the temporal moments---the model $\widehat{D^{01}}$ become lesser in magnitude relative to the truth as $A$ increases.
%
It is not immediately clear whether these differences are due to issues with self-similar convergence at high $A$ or they are indicating that the proposed model is lacking higher-order nonlocal information.
To assess this, the Atwood dependency of nonlocality proposed here should be incorporated into a full RANS model to be spatio-temporally evolved.
This should be the subject of future work.

\begin{table}[]
    \centering
    {\renewcommand{\arraystretch}{1.15}%
    \begin{tabular}{c l}
         $A$ &  $-\frac{\widehat{D^{01}}}{\widehat{D^{00}}}$ \\
         \hline
         $0.05$ & $0.11$\\
         $0.3$ & $0.13$\\
         $0.5$ & $0.16$\\
         $0.8$ & $0.16$
    \end{tabular}}
    \caption{Maximum ratio of magnitudes of first temporal moment and leading order moment from the proposed model for each Atwood case.}
    \label{tab:model_moment_ratios}
\end{table}

In Table \ref{tab:model_moment_ratios}, the $-\frac{\widehat{D^{01}}}{\widehat{D^{00}}}$ from the proposed model is presented to confirm that it goes below the threshold of $0.25$ for all Atwood cases.
For this model, $-\frac{\widehat{D^{01}}}{\widehat{D^{00}}}$ increases with $A$ but remains below $0.25$.

\section{Conclusion}
\label{sec:conclusion}

In this work, MFM is used to measure the eddy diffusivity moments associated with mean scalar transport in turbulent RT mixing for different Atwood numbers.
Similarly to a past work studying 2D, low-Atwood RT \citep{lavacot2024}, it is found here that nonlocality is important for  modeling 3D RT mixing.
There are several takeaways from this work:
\begin{enumerate}
\item Over the Atwood numbers studied here ($A=0.05$, $A=0.3$, $A=0.5$, and $A=0.8$), it is found that the importance of nonlocality increases as $A$ increases.
This is observed through examination of the eddy diffusivity moments measured using MFM and the terms of the Kramers-Moyal expansion.
Higher-order terms become closer in magnitude to the leading order moment with increasing $A$.
This suggests nonlocality is especially important in modeling RT mixing at higher $A$.
\item Temporal nonlocality appears to be more important than spatial nonlocality.
In testing different combinations of eddy diffusivity moments in an inverse operator, it is found that addition of temporal moments results in the most significant changes in predictions of $F$.
The predictions for the $A=0.05$ and $A=0.3$ cases are close to $F$ from high-fidelity simulations.
\item The eddy diffusivity moments must satisfy the constraint $-\frac{\widehat{D^{01}}}{\widehat{D^{00}}}<0.25$ in the self-similar zone for an inverse operator described by Equation \ref{eq:MMI_D00_D10_D01_D20} to be robust.
That is, inverse operators using moments that violate this constraint do not have dissipation terms, resulting in unstable solutions.
It appears that RT mixing that is not far into the self-similar regime does not satisfy this constraint.
\item Higher-order eddy diffusivity moments take longer to reach self-similarity than lower-order moments, and this effect is greater with increasing $A$.
This means that even if certain metrics for self-similarity (e.g., convergence of $\alpha$ or $\phi$) are met, higher-order eddy diffusivity moments may not yet be self-similar.
Thus, it is important to carefully examine the self-similarity of the higher-order moments themselves when making conclusions about the self-similarity of the turbulent mixing.
\item 
\edit{
	An inverse operator with algebraically-defined coefficients is proposed for mean scalar transport. 
	This is presented as a framework for incorporating non-locality and its dependence on Atwood number, with the goal of utilizing this to improve RANS models in future work.
}
\end{enumerate}

Througn examination of four Atwood number cases, an Atwood number dependence has been identified in the relative importance of nonlocal terms.
To further quantify the Atwood dependence, future work should perform these analyses at more Atwood numbers.
Particularly, it would be helpful to study intermediate $A$ between $0.05$ and $0.3$ to identify trends or transitions in behavior from low to intermediate Atwood numbers.
This would give better insight into the Atwood dependence of nonlocality and potentially allow for robust quantification of this dependence.

More work should be done to explore incorporation of the Atwood-dependence of nonlocality into turbulence models.
The model proposed in \S \ref{sec:atwood_model} uses fairly simple functions of $\eta$ for its coefficients, and these functions were chosen to fit a limited amount of data.
More Atwood cases should be studied to provide more data for tuning the coefficients to give more accurate predictions.
Additionally, an \textit{a priori} assessment of the model is presented here, in that quantities from the simulation, particularly $\rhom$ and $\YHm$, were used to solve $F$ transport equation.
A more thorough assessment would involve solving the full set of model equations including those for scalar transport, momentum, and density.

It is also found that the high Atwood simulations do not go far into the self-similar regime, and that the moments measured from these cases do not exhibit good self-similar collapse.
Data from later in the self-similar regime for these Atwood cases would improve this analysis.
This would likely require higher resolution simulations that allow the RT mixing to develop further into the self-similar regime.

While the present work is an analysis in the self-similar regime, future work should consider MFM analysis in spatio-temporal coordinates.
Ultimately, a model that accurately predicts mixing \edit{across all} regimes is desired, which would require departure from self-similar analysis.
MFM analysis in spatio-temporal coordinates and examination of the eddy diffusivity moments across the turbulent transition could facilitate progress towards this goal.

Overall, unique insights are acquired into the importance of nonlocality in variable density 3D RT mixing.
To make turbulent mixing models for RT instability more accurate, this nonlocality must be incorporated into the models.
The work presented here lays out the first steps towards constructing such a model based on direct measurements of the eddy diffusivity governing mean scalar transport. 
Future work will investigate practical ways to incorporate this information into RANS models, such as the $k$--$L$ model \cite{dimontetipton2006}.

\textbf{Acknowledgements.} 
This work was performed under the auspices of the US Department of Energy by Lawrence Livermore National Laboratory under Contract No. DE-AC52-07NA27344.

\appendix

\section{Artificial fluid method in \textsc{Pyranda}}
\label{appendix:artificial_fluid}
Artificial molecular viscosity, bulk viscosity, thermal conductivity, and species diffusivity are computed and added to the physical fluid properties to dampen numerical instabilities that may arise due to the high-order numerics:
\edit{
	\begin{align}
		\mu &= \mu_f + \mu^*,\\
		\beta &= \beta_f + \beta^*,\\
		\kappa &= \kappa_f + \kappa^*,\\
		D &= D_f +  D^*,\\
	\end{align}
	where $f$ denotes the physical fluid property, and $*$ denotes the artificial quantities, defined as follows:
	\begin{align}
		\mu^* &= 10^{-4}\rho\overline{|\mathcal{F}(S_{ij}S_{ij})|}\Delta^2,\\
		\beta^* &= 7\times10^{-2}\overline{\rho\left|\mathcal{F}\bigpar{\frac{\partial u_i}{\partial x_i}}\right|}\Delta^2,\\
		\kappa^* &= 10^{-3}\frac{\overline{\rho c_v \mathcal{F}(T)}}{T}\frac{\Delta^2}{\Delta t},\\
		D^* &=\max\bigbra{
			10^{-4}\overline{\mathcal{F}(Y_H)},
			10^2\overline{|Y_H|-1+|1-Y_H|}
		}\frac{\Delta^2}{\Delta t}.\\
	\end{align}
	Above, the bar is the Gaussian filter described in \citet{cook2007}, and $S_{ij}$ is the strain rate tensor, defined as $\frac{1}{2}\bigpar{\frac{\partial u_i}{\partial x_j}+\frac{\partial u_j}{\partial x_i}}$.
	$\mathcal{F}$ is the eighth-order operator:
	\begin{align}
		\mathcal{F}=\max\bigpar{\frac{\partial^8}{\partial x^8}\Delta^8,\frac{\partial^8}{\partial y^8}\Delta^8,\frac{\partial^8}{\partial z^8}\Delta^8}.
	\end{align}
}

\section{Autocorrelation Curves in $y$ from 3-D Simulations}
\label{appendix:autocorr_y}

\begin{figure}[t]
	\centering
	\begin{subfigure}[]{0.3\textwidth}
		\includegraphics[width=\textwidth]{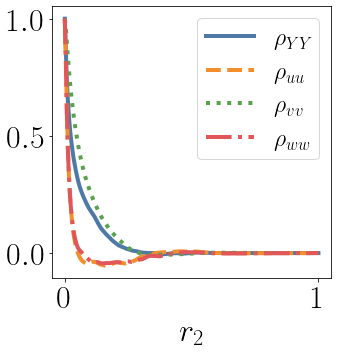}
		\subcaption[]{$A=0.05$}
	\end{subfigure}
	\begin{subfigure}[]{0.3\textwidth}
		\includegraphics[width=\textwidth]{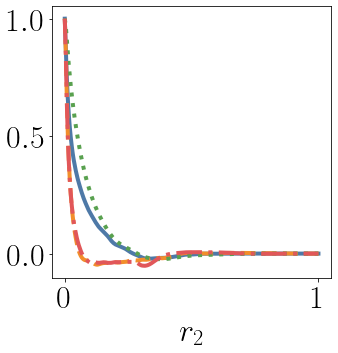}
		\subcaption[]{$A=0.5$}
	\end{subfigure}
	\begin{subfigure}[]{0.3\textwidth}
		\includegraphics[width=\textwidth]{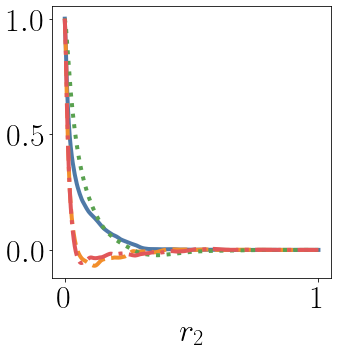}
		\subcaption[]{$A=0.8$}
	\end{subfigure}
	\caption{Correlation curves over $r_2$ at different Atwood numbers at the last timesteps of the simulations.
		Curves are averaged over $x$ and $z$.}
	\label{fig:Rii_y}
	\vskip 1em 
	\centering
	\begin{subfigure}[]{0.3\textwidth}
		\includegraphics[height=11em]{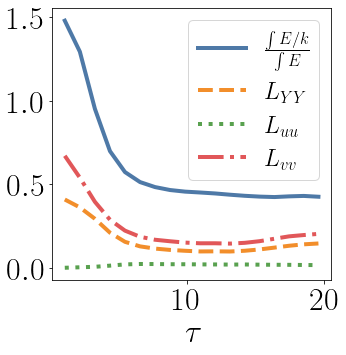}
		\subcaption[]{$A=0.05$}
	\end{subfigure}
	\begin{subfigure}[]{0.3\textwidth}
		\includegraphics[height=11em]{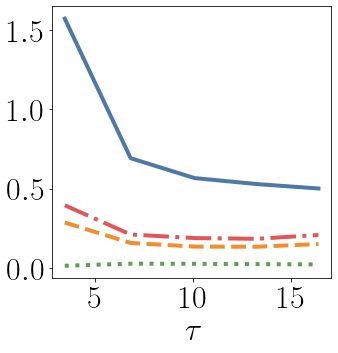}
		\subcaption[]{$A=0.5$}
	\end{subfigure}
	\begin{subfigure}[]{0.3\textwidth}
		\includegraphics[height=11em]{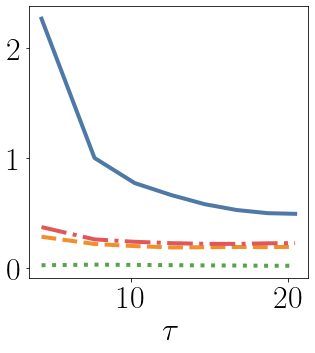}
		\subcaption[]{$A=0.8$}
	\end{subfigure}
	\caption{Integral length scales at different Atwood numbers over time using Equation \ref{eq:LYY_2} (solid blue) and Equation \ref{eq:LYY_1} for $Y_H''$ (dashed orange), $u''$ (dotted green), and $v''$ (dash-dotted red).}
	\label{fig:L_y}
\end{figure}

\edit{
	The autocorrelation in $y$ is defined as
	\begin{align}
		\rho_{qq}(r_2) = \frac{\langle q(x,y,z,t)q(x,y+r_2,z,t)\rangle}{\langle q(x,y,z,t)^2\rangle},
	\end{align}
	which we take over the entire domain in $y$.
	Here, $\langle * \rangle$ denotes averaging over $x$ and $z$ for one realization.
	The autocorrelations are plotted in \ref{fig:Rii_y}, which we observe also decay to zero.
	Based on $\rho_{qq}(r_2)$, we define the normalized integral length scale:
	\begin{align}
		L_{qq}=\frac{\int\rho_{YY}(r_2)dr_2}{h_b}.\label{eq:LYY_1}
	\end{align}
	$L_I$ can also be approximated using the energy spectrum, as shown by \citet{morgan2017}:
	\begin{align}
		L_I = \frac{1}{h_b}\frac{\int k^{-1}E(k) dk}{\int E(k) dk}. \label{eq:LYY_2}
	\end{align}
}
\edit{
	These length scales are plotted in Figure \ref{fig:L_y}.
	$L_{YY}$ and $L_{vv}$, which are based on quantities dominated by the gravitational acceleration, are similar and larger than $L_{uu}$ over all $\tau$.
	In late time, $L_I$ is nearly parallel with $L_{YY}$ and $L{vv}$.
	At early times, $L_I$ deviates more from $L_{YY}$ and $L{vv}$, and this mismatch becomes larger as $A$ increases.
}

\section{Self-similar collapse of eddy diffusivity moments}

\begin{figure}[H]
    \centering
        \includegraphics[width=\textwidth]{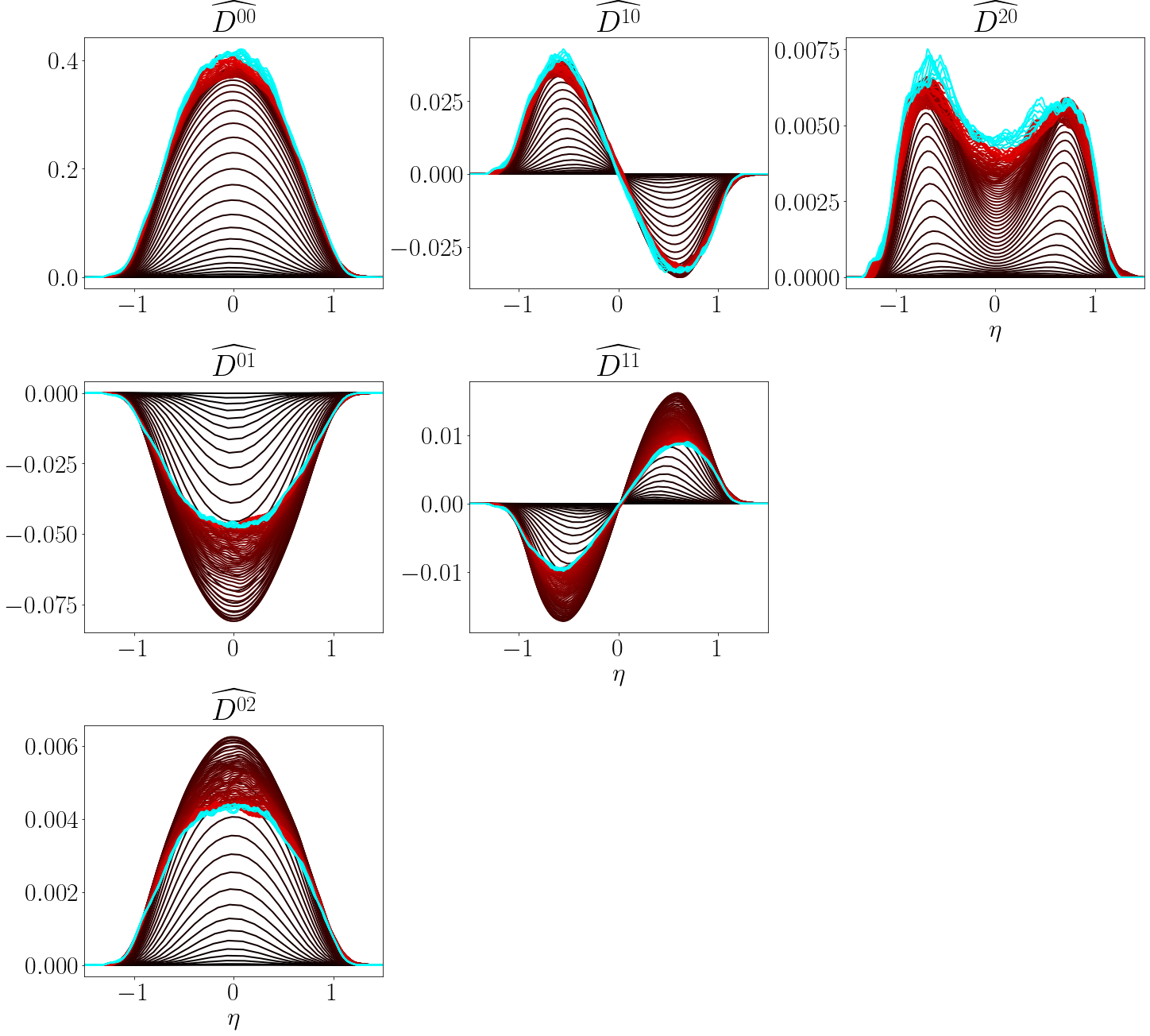}
        \caption{Self-similar collapse of eddy diffusivity moments at $A=0.05$. Dark lines are earlier times, and light lines are later times.}
    \label{fig:selfsim_A005}
\end{figure}

\begin{figure}[H]
    \centering
        \includegraphics[width=\textwidth]{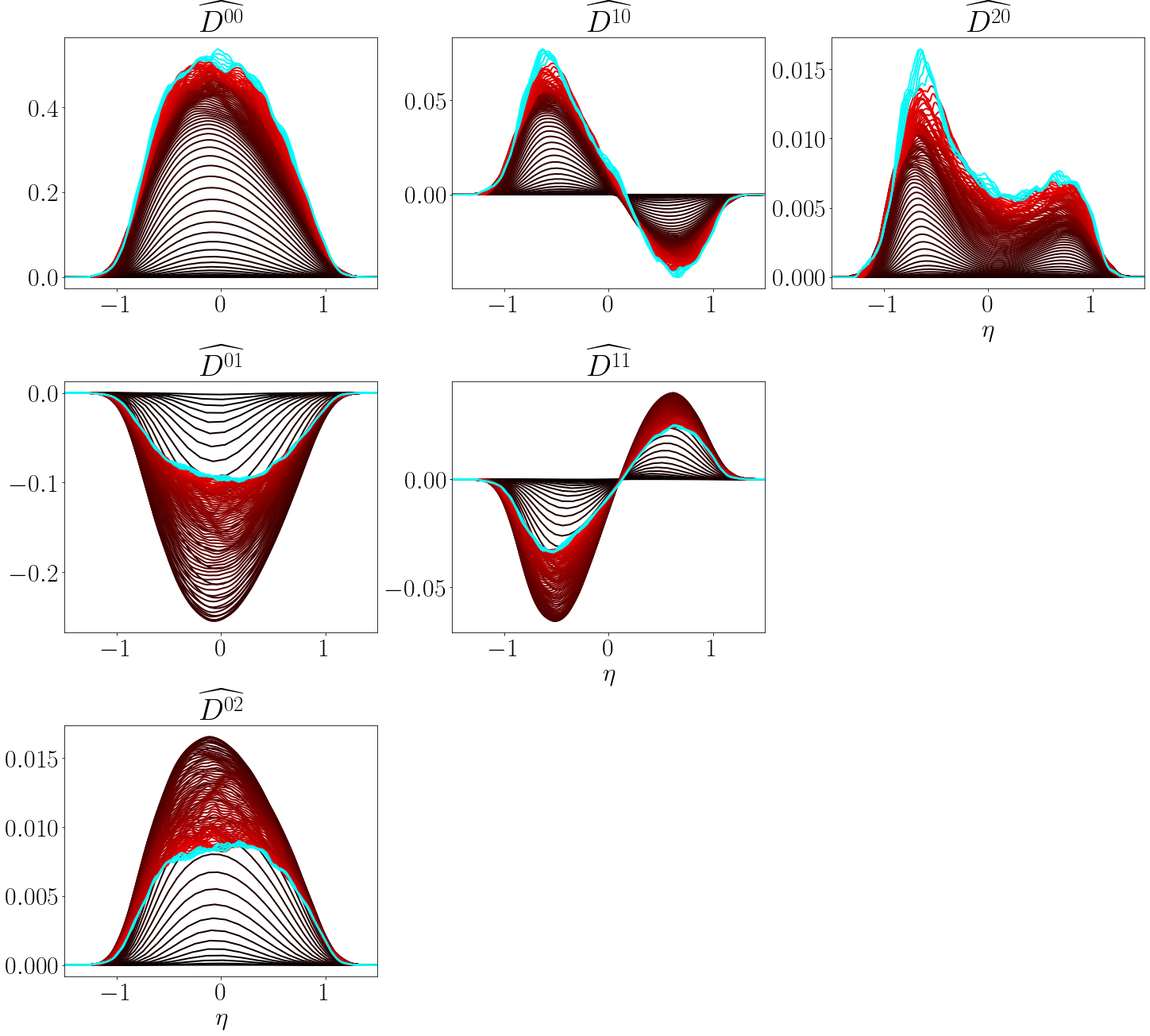}
        \caption{Self-similar collapse of eddy diffusivity moments at $A=0.3$. Dark lines are earlier times, and light lines are later times.}
    \label{fig:selfsim_A03}
\end{figure}

\begin{figure}[H]
    \centering
        \includegraphics[width=\textwidth]{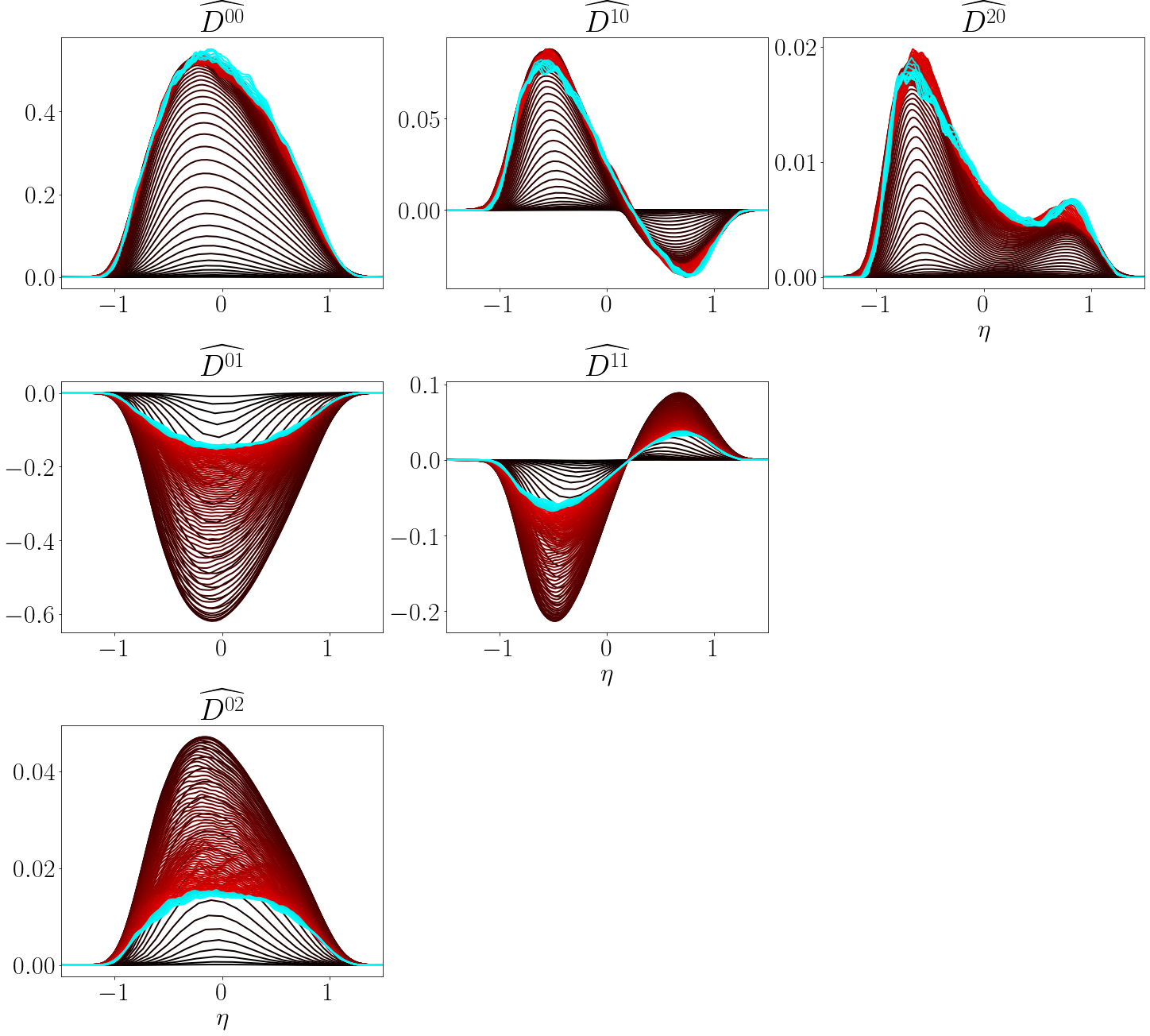}
        \caption{Self-similar collapse of eddy diffusivity moments at $A=0.5$. Dark lines are earlier times, and light lines are later times.}
    \label{fig:selfsim_A05}
\end{figure}

\begin{figure}[H]
    \centering
        \includegraphics[width=\textwidth]{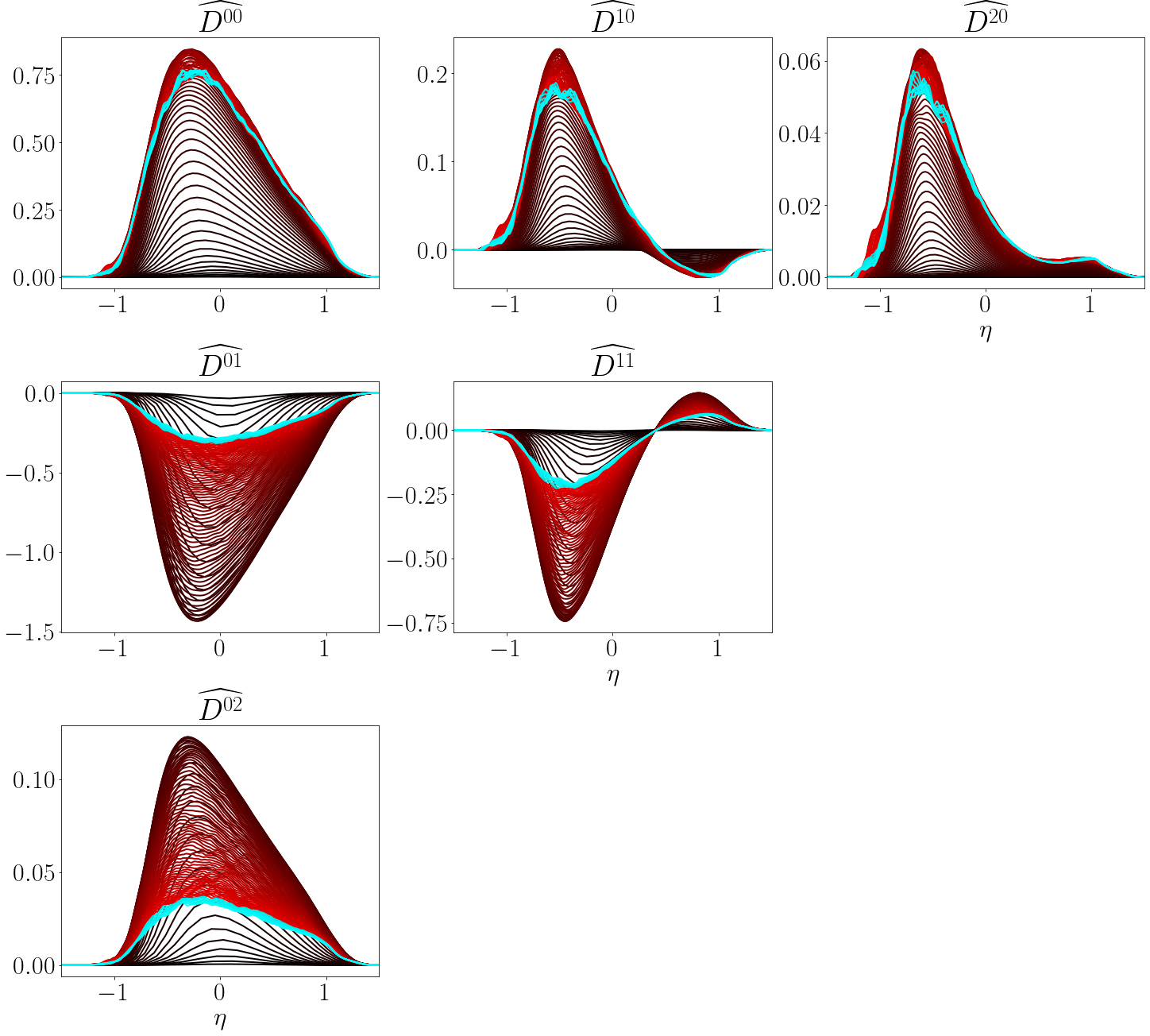}
        \caption{Self-similar collapse of eddy diffusivity moments at $A=0.8$. Dark lines are earlier times, and light lines are later times.}
    \label{fig:selfsim_A08}
\end{figure}

Figures \ref{fig:selfsim_A005} -  \ref{fig:selfsim_A08} show the self-similar collapse of the eddy diffusivity moments.
Normalization is applied according to Equations \ref{eq:selfsim1} - \ref{eq:selfsim2}.

\section{Requirements on eddy diffusivity moments for a robust model form}
\label{sec:model_robustness}
For simplicity and ease of algebra, we derive the requirements on eddy diffusivity moments for a robust inverse operator using the moments $D^{00}$, $D^{10}$, and $D^{01}$.
It has been found that the same requirement on $-\frac{D^{10}}{D^{00}}$  is recovered for inverse operators using $D^{00}$, $D^{10}$, $D^{01}$, and  $D^{20}$ through a similar analysis.

The inverse operator using $D^{00}$, $D^{10}$, and $D^{01}$ is
\begin{align}
	\bigbra{1 + a^{10}\diffp{}{y} + a^{01}\diffp{}{t} }F = a^{00}\rhom\diffp{\YHm}{y}.
	\label{eq:MMI_D00_D10_D01}
\end{align}

\begin{figure}
	\centering
	\begin{subfigure}[]{0.3\textwidth}
		\includegraphics[height=12em]{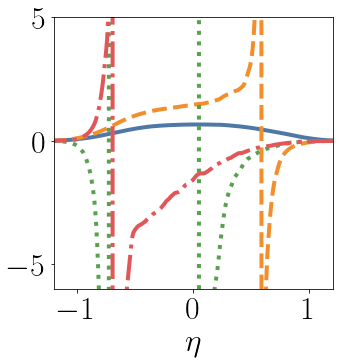}
		\subcaption[]{$a^{00}$}
		\label{subfig:MMI_coeffs_moms_012_0}
	\end{subfigure}
	\begin{subfigure}[]{0.3\textwidth}
		\includegraphics[height=12em]{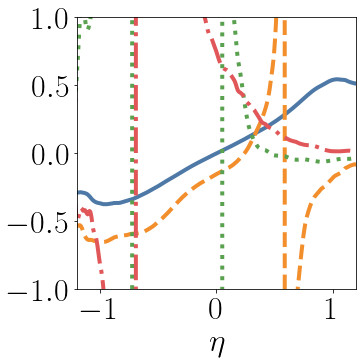}
		\subcaption[]{$a^{10}$}
		\label{subfig:MMI_coeffs_moms_012_1}
	\end{subfigure}
	\begin{subfigure}[]{0.3\textwidth}
		\includegraphics[height=12em]{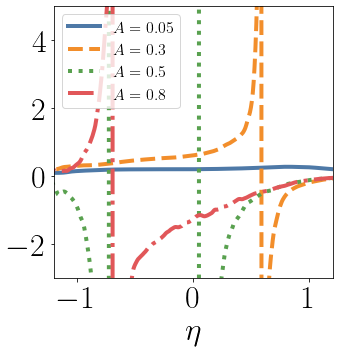}
		\subcaption[]{$a^{01}$}
		\label{subfig:MMI_coeffs_moms_012_2}
	\end{subfigure}
	\caption{MMI coefficients determined using $D^{00}$, $D^{10}$, and $D^{01}$ for different Atwood numbers. }
	\label{fig:D00_D01_coeffs}
\end{figure}

Figure \ref{fig:D00_D01_coeffs} shows the MMI coefficients $a^{mn}$ for the above model, determined from the MFM-measured eddy diffusivity moments for each Atwood case.
The two lowest Atwood cases, $A=0.05$ and $A=0.3$, have positive $a^{01}$.
It appears that for higher Atwood numbers, the MMI process results in negative $a^{01}$ and, therefore, models with no destruction terms.
This indicates that 1) the MMI process may not always produce a robust model, depending on the form of the inverse operator, and 2) there must be some crossover $A$ over which the model that results from the MMI process changes from robust to non-robust.

Based on the observation that, with certain combinations of eddy diffusivity moments, MMI does not produce robust models for higher Atwood numbers, there must be a requirement on eddy diffusivity moments for MMI to result in a robust model, depending on the form of the inverse operator.
To determine this requirement for different operators, the MMI equations (Equations \ref{eq:MMI_num_1} - \ref{eq:MMI_num_2}) are analytically solved in self-similar space and obtain the MMI coefficients in terms of the eddy diffusivity moments used in the models and their derivatives.

The self-similar MMI equations for the operator in Equation \ref{eq:MMI_D00_D10_D01} are
\begin{align}
	\widehat{a^{00}}\rhom - \widehat{a^{10}}\diff{}{\eta} \widehat{F^{00}} - \widehat{a^{01}}\bigpar{3-2\eta\diff{}{\eta}}\widehat{F^{00}} &= \widehat{F^{00}}, \\
	\widehat{a^{00}}\rhom\eta - \widehat{a^{10}}\diff{}{\eta} \widehat{F^{10}} - \widehat{a^{01}}\bigpar{5-2\eta\diff{}{\eta}}\widehat{F^{10}}&= \widehat{F^{10}},    \\
	\widehat{a^{00}}\rhom- \widehat{a^{10}}\diff{}{\eta} \widehat{F^{01}}  - \widehat{a^{01}}\bigpar{4-2\eta\diff{}{\eta}}\widehat{F^{01}} &= \widehat{F^{01}}.
\end{align}
An MMI fitting matrix $M$ is constructed from the terms on the left-hand sides of these equations.
The determinant of that matrix is
\begin{align}
	\mathcal{D} &= \det(M) = \rhom^3\bigpar{4\widehat{D^{01}} \widehat{D^{10}}^\prime + \widehat{D^{00}} \widehat{D^{10}}^\prime + 4\widehat{D^{01}} \widehat{D^{00}} + \widehat{D^{00}}^2 - 2\eta \widehat{D^{01}}^\prime \widehat{D^{00}} - 5\widehat{D^{10}} \widehat{D^{01}}^\prime}.
\end{align}
The MMI coefficents are then
\begin{align}
	\widehat{a^{00}} &= \frac{\mathcal{D}^{00}}{\mathcal{D}}, \quad
	\widehat{a^{10}} = \frac{\mathcal{D}^{10}}{\mathcal{D}}, \quad
	\widehat{a^{01}} = \frac{\mathcal{D}^{01}}{\mathcal{D}},
\end{align}
where
\begin{align}
	\mathcal{D}^{00} =  &\rhom^3 \left(-2\eta {\widehat{D^{00}}}^2 \widehat{D^{01}}^{\prime} + 2\eta \widehat{D^{00}} \widehat{D^{01}} \widehat{D^{00}}^{\prime} + {\widehat{D^{00}}}^3 + \widehat{D^{00}}{}^2 \widehat{D^{01}} + {\widehat{D^{00}}}^2 \widehat{D^{10}}^{\prime} + \widehat{D^{00}} \widehat{D^{01}} \widehat{D^{10}}^{\prime} \right. \nonumber\\
	&\quad \left. - \widehat{D^{00}} \widehat{D^{10}} \widehat{D^{00}}^{\prime} - 2 \widehat{D^{00}} \widehat{D^{10}} \widehat{D^{01}}^{\prime} + \widehat{D^{01}} \widehat{D^{10}} \widehat{D^{00}}^{\prime} \right),\\
	\mathcal{D}^{10} = &\rhom^3 \left(-2\eta \widehat{D^{01}} \widehat{D^{10}}^{\prime} + 2\eta \widehat{D^{10}} \widehat{D^{01}}^{\prime} - \widehat{D^{00}} \widehat{D^{10}} + \widehat{D^{01}} \widehat{D^{10}} \right),\\
	\mathcal{D}^{01} = 
	&\rhom^3 \left(-\widehat{D^{00}} \widehat{D^{01}} - \widehat{D^{01}} \widehat{D^{10}}^{\prime} + \widehat{D^{10}} \widehat{D^{01}}^{\prime} \right).
\end{align}

\begin{figure}
	\centering
	\begin{subfigure}[]{0.24\textwidth}
		\includegraphics[height=9em]{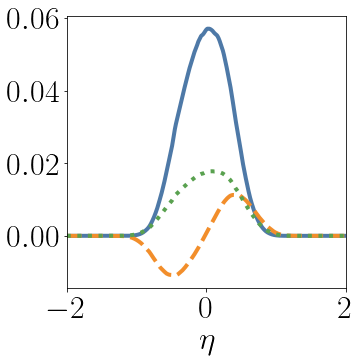}
	\subcaption[]{$A=0.05$}
	\label{subfig:nums_A005}
\end{subfigure}
\begin{subfigure}[]{0.24\textwidth}
	\includegraphics[height=9em]{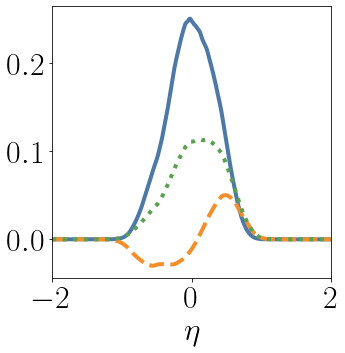}
\subcaption[]{$A=0.3$}
\label{subfig:numsA03}
\end{subfigure}
\begin{subfigure}[]{0.24\textwidth}
\includegraphics[height=9em]{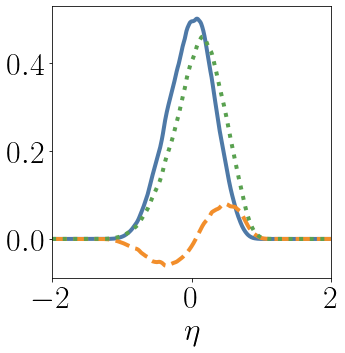}
\subcaption[]{$A=0.5$}
\label{subfig:nums_A05}
\end{subfigure}
\begin{subfigure}[]{0.24\textwidth}
\includegraphics[height=9em]{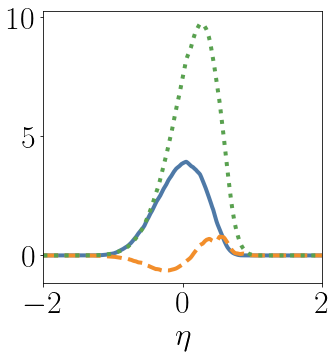}
\subcaption[]{$A=0.8$}
\label{subfig:nums_A08}
\end{subfigure}
\caption{Numerators of the MMI coefficients for the moment combination $D^{00}$, $D^{01}$, and $D^{10}$ for each $A$.
Solid blue: $a^{00}$, dashed orange: $a^{10}$, dotted green: $a^{01}$.}
\label{fig:nums}
\end{figure}

\begin{figure}
\centering
\begin{subfigure}[]{0.24\textwidth}
\includegraphics[height=9em]{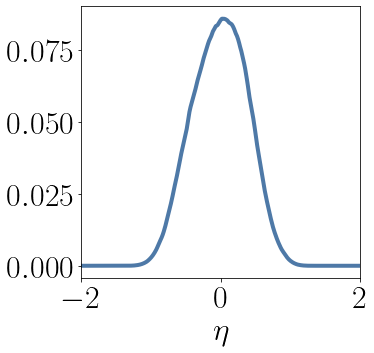}
\subcaption[]{$A=0.05$}
\label{subfig:dets_A005}
\end{subfigure}
\begin{subfigure}[]{0.24\textwidth}
\includegraphics[height=9em]{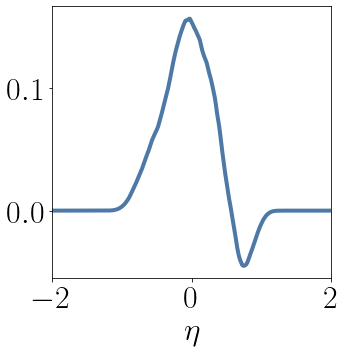}
\subcaption[]{$A=0.3$}
\label{subfig:detsA03}
\end{subfigure}
\begin{subfigure}[]{0.24\textwidth}
\includegraphics[height=9em]{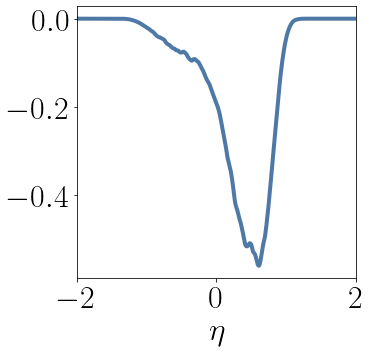}
\subcaption[]{$A=0.5$}
\label{subfig:dets_A05}
\end{subfigure}
\begin{subfigure}[]{0.24\textwidth}
\includegraphics[height=9em]{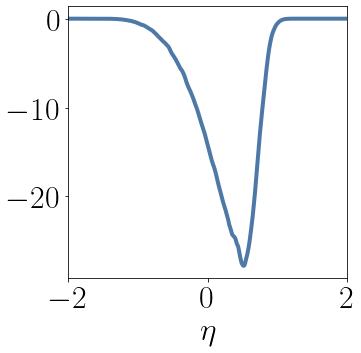}
\subcaption[]{$A=0.8$}
\label{subfig:dets_A08}
\end{subfigure}
\caption{Determinants of the MMI fitting matrix for the moment combination $D^{00}$, $D^{01}$, and $D^{10}$ for each $A$.}
\label{fig:dets}
\end{figure}

Concerning the robustness of this inverse operator, the relevant coefficient is $\widehat{a^{01}}$.
As seen in Figure \ref{fig:nums}, over the $A$ studied, the numerator of this coefficient does not change sign, so whether or not the coefficient changes sign is determined by the determinant of the MMI fitting matrix.
Figure \ref{fig:dets} shows the determinants of the MMI matrix. 
At $A=0.05$, the determinant is always positive, but flips sign between $A=0.3$ and $A=0.5$.
Additionally, $\mathcal{D}$ for $A=0.3$ crosses zero for some $\eta$.
Since eddy diffusivity moments from the $A=0.05$ case (Figure \ref{fig:D00_D01_coeffs}) gives MMI coefficients that have the correct signs, $\mathcal{D}$ must be positive for this operator (using $D^{00}$, $D^{01}$, and $D^{10}$) to be robust.

\begin{figure}
\centering
\begin{subfigure}[]{0.24\textwidth}
\includegraphics[height=9em]{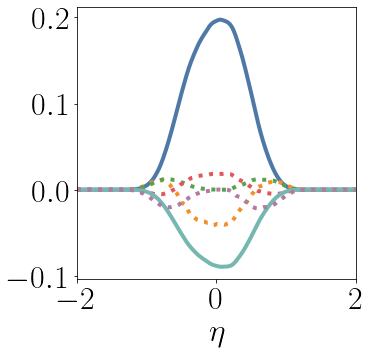}
\subcaption[]{$A=0.05$}
\label{subfig:detterms_A005}
\end{subfigure}
\begin{subfigure}[]{0.24\textwidth}
\includegraphics[height=9em]{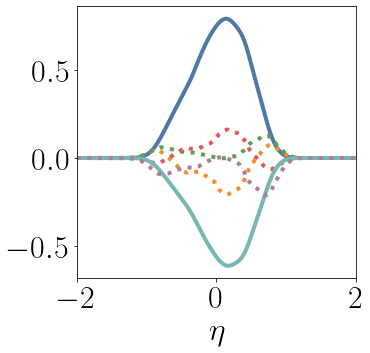}
\subcaption[]{$A=0.3$}
\label{subfig:dettermsA03}
\end{subfigure}
\begin{subfigure}[]{0.24\textwidth}
\includegraphics[height=9em]{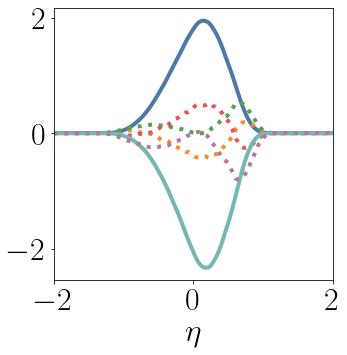}
\subcaption[]{$A=0.5$}
\label{subfig:detterms_A05}
\end{subfigure}
\begin{subfigure}[]{0.24\textwidth}
\includegraphics[height=9em]{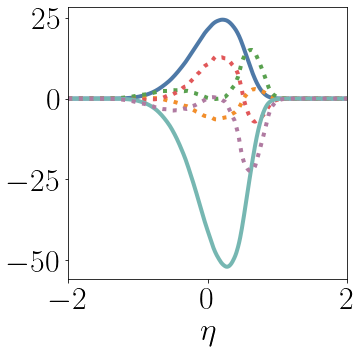}
\subcaption[]{$A=0.8$}
\label{subfig:detterms_A08}
\end{subfigure}
\caption{Terms composing the determinants of the MMI fitting matrix for the moment combination $D^{00}$, $D^{01}$, and $D^{10}$ for each $A$.
Solid lines are the terms $4\widehat{D^{01}}\widehat{D^{00}}$ (teal) and $\widehat{D^{{00}}}^2$ (blue).}
\label{fig:detterms}
\end{figure}

To determine the requirements on the eddy diffusivity moments for $\mathcal{D}>0$,the terms that compose $\mathcal{D}$, which are plotted in Figure \ref{fig:detterms}, are examined.
Based on those plots, $4\widehat{D^{01}}\widehat{D^{00}}$ and $\widehat{D^{{00}}}^2$ are the dominant terms in the determinant.
Thus, when $-\frac{\widehat{D^{01}}}{\widehat{D^{00}}}\approx0.25$, the determinant becomes zero, and there are no solutions for the coefficients.
For $-\frac{\widehat{D^{01}}}{\widehat{D^{00}}}<0.25$, the model coefficients have the correct sign; at $A=0.05$, $-\frac{\widehat{D^{01}}}{\widehat{D^{00}}}\approx0.1$.
For $-\frac{\widehat{D^{01}}}{\widehat{D^{00}}}>0.25$, the model coefficients flip sign; at $A=0.5$, $-\frac{\widehat{D^{01}}}{\widehat{D^{00}}}\approx0.4$.
Thus, for a robust first spatio-temporal inverse operator, $-\frac{\widehat{D^{01}}}{\widehat{D^{00}}}$ must be under $0.25$.

\bibliographystyle{bibsty}
\bibliography{bibliography.bib}

\end{document}